\pdfoutput=1
\documentclass{elsarticle}
\usepackage{graphicx}
\usepackage{psfrag}
\usepackage{makeidx}  
\usepackage{multirow}
\usepackage{psfrag}

\begin{document}

\title{A Signed Particle Formulation of Non-Relativistic Quantum Mechanics}

\author[bg]{J.M.~Sellier$^*$}
\address[bg]{IICT, Bulgarian Academy of Sciences, Acad. G.~Bonchev str. 25A, 1113 Sofia, Bulgaria\\$^*$\texttt{jeanmichel.sellier@parallel.bas.bg,\\jeanmichel.sellier@gmail.com}}

\begin{abstract}
A formulation of non-relativistic quantum mechanics in terms of Newtonian particles is presented
in the shape of a set of three postulates.
In this new theory, quantum systems are described by ensembles of signed particles which behave
as field-less classical objects which carry a negative or positive sign and interact with
an external potential by means of creation and annihilation events only.
This approach is shown to be a generalization of the signed particle Wigner Monte Carlo method which reconstructs
the time-dependent Wigner quasi-distribution function of a system
and, therefore, the corresponding Schr\"{o}dinger time-dependent wave-function.
Its classical limit is discussed and a physical interpretation, based on experimental evidences coming from quantum tomography, is suggested.
Moreover, in order to show the advantages brought by this novel formulation, a straightforward extension to relativistic effects is discussed.
To conclude, quantum tunnelling numerical experiments are performed to show the validity of the suggested approach.
\end{abstract}


\maketitle

\section{Introduction}

Nowadays, many different mathematical formulations of quantum mechanics exist,
among which the ones suggested by E.~Schr\"{o}dinger \cite{Schrodinger},
E.~Wigner \cite{Wigner},
R.~Feynman \cite{Feynman},
L.V.~Keldysh \cite{Keldysh},
K.~Husimi \cite{Husimi},
D.~Bohm \cite{Bohm01}, \cite{Bohm02} are the most popular ones.
While, at a first glance they seem to be drastically different theories, it can be shown that they are all mathematically equivalent.
As a matter of fact, they offer the same set of predictions and can be seen as complementary points of view.
The situation is not any different than what one observes in classical mechanics where different
formalisms, such as Newtonian, Lagrangian, Hamiltonian, etc., coexist and shed different light on different
mechanical aspects.

In this paper, we introduce a new formulation of quantum mechanics in terms of signed classical field-less particles.
This suggested theory is based on a generalization and a novel physical interpretation of the mathematical Wigner Monte Carlo method \cite{MCMA}
which is able to simulate the time-dependent {\sl{single}}- and {\sl{many}}-body Wigner equation \cite{JCP-01}, \cite{JCP-02}
in a quite intuitive fashion, which experimentalists are familiar with. Indeed it describes quantum objects in terms
of classical particles only.

One should notice that the signed particle formulation is equivalent to the usual formulations.
As such, no fundamental new results are introduced. The predictions made
are the same as the ones made in the more standard theory.
However, {\sl{"there is a pleasure in recognizing old things from a new point of view"}} \cite{Feynman}
and the author hopes it can offer a new perspective on the puzzling quantum nature of the microscopic world.
The new theory is based on classical particles which have a position and momentum simultaneously
although the Heisenberg principle of uncertainty is still respected in the formulation and embedded in terms of unknown
initial conditions. In particular, we will show that the sign of a particle cannot be evaluated experimentally
and no physical measurement can be depicted to find differences with other formalisms.
Nevertheless, it offers some noticeable advantages.
On the one hand it is a very intuitive approach which provides a new way to describe Nature at a quantum level.
On the other hand, it is a computationally attractive formulation
being based on independently evolving particles, allowing deep levels of parallelization in the time-dependent simulation of
quantum single- and many-body systems.
Finally, it allows the inclusion of physical effects which are difficult to treat in other formulations of quantum mechanics.

In this paper, we first introduce the three postulates which completely define the new mathematical formulation of quantum mechanics in terms of
signed particles. We then proceed with showing that these three postulates are enough to replicate the results
of more conventional quantum theories. In particular we show that our suggested approach
is a natural generalization of the Wigner Monte Carlo method which
reconstructs the time-dependent Wigner quasi-distribution function and, thus, the corresponding Schr\"{o}dinger wave-function.
Afterwards, its classical limit is considered in several details.
In order to show the applicability of the proposed theory, we numerically simulate several experiments involving quantum tunnelling.
To conclude, we extend the theory to include general relativity in the formalism by generalizing the second postulate
to the case of particles travelling along space-time geodesics, and we show that typical relativistic effects are
observable for quantum wave-packets.

\section{A signed particle approach}

We now introduce the signed particle formulation of quantum mechanics by enunciating, and discussing, the three postulates
which completely define the new theory.

\bigskip

{\sl{{\bf{Postulate I.}} Physical systems can be described by means of (virtual) Newtonian particles, i.e. provided with a position ${\bf{x}}$ and a momentum ${\bf{p}}$ simultaneously, which carry a sign which can be positive or negative.}}

\bigskip

{\sl{{\bf{Postulate II.}} A signed particle, evolving in a potential $V=V \left( {\bf{x}} \right)$, behaves as a
field-less classical point-particle which, during the time interval $dt$, creates a new pair of signed particles
with a probability $\gamma \left( {\bf{x}}(t) \right) dt$ where
\begin{equation}
 \gamma\left( {\bf{x}} \right) = \int_{-\infty}^{+\infty} \mathcal{D}{\bf{p}}' V_W^+ \left( {\bf{x}}; {\bf{p}}' \right)
\equiv \lim_{\Delta {\bf{p}}' \rightarrow 0^+} \sum_{{\bf{M}} = -\infty}^{+\infty} V_W^+ \left( {\bf{x}}; {\bf{M}} \Delta {\bf{p}}' \right),
\label{momentum_integral}
\end{equation}
and $V_W^+ \left( {\bf{x}}; {\bf{p}} \right)$ is the positive part of the quantity
\begin{equation}
 V_W \left( {\bf{x}}; {\bf{p}} \right) = \frac{i}{\pi^d \hbar^{d+1}} \int_{-\infty}^{+\infty} d{\bf{x}}' e^{-\frac{2i}{\hbar} {\bf{x}}' \cdot {\bf{p}}} \left[ V \left( {\bf{x}}+{\bf{x}}'\right) - V \left( {\bf{x}}-{\bf{x}}'\right)  \right],
\label{wigner-kernel}
\end{equation}
known as the Wigner kernel (in a $d$-dimensional space) \cite{Wigner}. If, at the moment of creation, the parent particle has sign $s$,
position ${\bf{x}}$ and momentum ${\bf{p}}$,
the new particles are both located in ${\bf{x}}$, have signs $+s$ and $-s$, and momenta ${\bf{p}}+{\bf{p}}'$ and ${\bf{p}}-{\bf{p}}'$ respectively,
with ${\bf{p}}'$ chosen randomly according to the (normalized) probability $\frac{V_W^+ \left( {\bf{x}}; {\bf{p}} \right)}{\gamma({\bf{x}})}$.}}

\bigskip

{\sl{{\bf{Postulate III.}} Two particles with opposite sign and same phase-space coordinates $\left( {\bf{x}}, {\bf{p}}\right)$ annihilate.}}

\bigskip

We will show that this set of three postulates is sufficient to reconstruct
the time-dependent quasi-distribution function of a system and, consequentely, its
wave-function. In other words, the time-dependent evolution of a quantum system
can be {\sl{completely}} expressed in terms of creation and annihilation of signed particles only.
Some comments on the interpretation and application of the postulates follow below.

\bigskip

{\sl{Momentum integral}}. It is important to note that the definition of the function $\gamma = \gamma \left( {\bf{x}} \right)$
is introduced in a continuum phase-space which, in turn, introduces a non-Riemann integral (\ref{momentum_integral}).
This integral is, in the spirit, very similar to a path integral and the author proposes to name it a {\sl{momentum}} integral.
In fact, while a path integral is spanning the space of trajectories, the integral (\ref{momentum_integral}) similarly
spans the space of momenta. Of course, this new definition is rather elementary and necessitates more mathematical investigations
as it may not always imply convergence.

\bigskip

{\sl{Particles with a sign}}. The concept of signed particles was suggested for the first time in \cite{Ned04} to explain the emergence
of negative peaks in the Wigner quasi-distribution of particular quantum systems (and the reader should note that it was introduced,
strictly speaking, as a {\sl{mathematical}} tool).
Consecutively, this concept has been applied with success to the numerical simulation of a challenging
quantum benchmark test involving the evolution of $\delta-$functions \cite{APL-2013}.
Thus, we borrow this notion to enunciate the first postulate.
In particular, positive and negative particles can be interpreted as usual classical field-less particles moving on Newtonian trajectories,
but it is impossible to experimentally know simultaneously their position and momentum due to the Heisenberg
uncertainty principle, thus denying the chance of calculating their trajectory (and one can only start from a distribution of particles
in practical calculations).
The physical interpretation of the negative sign is suggested below based on experimental evidences
coming from the field of quantum tomography \cite{Leonhardt}, \cite{Monroe}, a set of
techniques to reconstruct the Wigner function of a given experimental setting.

\bigskip

{\sl{Negatively signed particles}}. In practice, an experiment is prepared in specific initial conditions and repeated a large amount of times
(quantum systems have different outcomes every time), providing a projection of
the corresponding quasi-distribution on the spatial coordinate axis and momentum axis,
and by applying the inverse Radon transformation one reconstructs a higher dimensional
function which is the Wigner function of the system (with positive and negative values).
Now, classical objects are always localized in a precise point of the phase-space (represented by $\delta-$functions), 
while for quantum objects the presence of the Heisenberg principle of uncertainty
prevents such localization, forcing the description of the dynamics to an area of
the phase-space bigger than $\Delta x \Delta p = \frac{\hbar}{2}$.
This is clearly exhibited by the appearance of negative values in the Wigner function and,
therefore, one may infer that particles with a negative sign
are essentially those which are experimentally unreachable \cite{Monroe}.

\bigskip

{\sl{Physical picture}}. Consequently, the picture that this novel theory offers is rather peculiar and different than any other
mathematical formulation of quantum mechanics. Quantum systems are now described by means of ensembles of Newtonian field-less particles
which now carry a sign and interact with an external potential
by means of creation and annihilation events only. When a pair of particles is created, one is in an experimentally reachable state (positive sign),
and the other in a non-reachable one (negative sign). This offers a physical picture which is relatively easy to grasp and which
allows the inclusion of quite complex effects in a natural way (as it will be shown below).

\bigskip


{\sl{Practical implementation of the formulation}}. In particular, given a signed particle at time $t$ with sign $s$, mass $m$, and (phase-space) coordinates $\left( {\bf{x}}, {\bf{p}}\right)$,
indicated from now on by $ \left( s, m; {\bf{x}}, {\bf{p}} \right)$,
we introduce the operator $\hat{S}$ which constructs a new set of {\sl{three}} signed particles
$$
 \hat{S} \left[ \left( s, m; {\bf{x}}, {\bf{p}} \right)  \right] = \left\{ \left( s, m; {\bf{x}}_1, {\bf{p}}_1 \right), \left( +s, m; {\bf{x}}_2, {\bf{p}}_2 \right), \left( -s, m; {\bf{x}}_3, {\bf{p}}_3 \right) \right\}
$$
in the following way:
\begin{itemize}
 \item
At time $t$, one generates a random number $r \in [0,1]$ and computes the quantity $\delta t=-\frac{\ln{r}}{\gamma \left( {\bf{x}}(t)  \right)}$,
 \item
At time $t+\delta t$, the initial particle evolves as field-less and has new coordinates $\left( {\bf{x}}_1, {\bf{p}}_1 \right) = \left( {\bf{x}}+\frac{{\bf{p}}(t)}{m} \delta t, {\bf{p}}  \right)$,
 \item
A pair of new signed particles is created at time $t+\delta t$, where the particle with the sign $s$ has coordinates
$\left( {\bf{x}}_2, {\bf{p}}_2 \right) = \left( {\bf{x}}+\frac{{\bf{p}}(t)}{m} \delta t, {\bf{p}}+{\bf{p}}'  \right) $ and the particle $-s$ has coordinates
$\left( {\bf{x}}_3, {\bf{p}}_3 \right) = \left( {\bf{x}}+\frac{{\bf{p}}(t)}{m} \delta t, {\bf{p}}-{\bf{p}}'  \right) $
and the quantity ${\bf{p}}'$ is computed from the normalized probability ${}^{\left| V_W \left({\bf{x}}; {\bf{p}} \right) \right|} /_{\gamma \left({\bf{x}}\right)}$.
\end{itemize}

The reason for generating a random number $r$ is easily explained. The probability that a signed particle, in a small interval $dt$,
generates a new pair of particles is $\gamma\left( {\bf{x}}(t) \right) dt$. Consequently, the probability that a particle
which has generated a new signed pair at time $t_0$ has not yet generated another new pair at time $t$ is
$$
 e^{-\int_{t_0}^{t} dt' \gamma\left( {\bf{x}}(t') \right)},
$$
or, in other words, the probability that no creation occurs during the interval $[t_0,t]$.
Therefore, the probability $P(t)$ that a particle will generate a new pair during the interval $dt$ is given by
$$
 P(t) dt = \gamma\left( {\bf{x}}(t) \right) e^{-\int_{t_0}^{t} dt' \gamma\left( {\bf{x}}(t') \right)} dt.
$$
Due to the complexity of evaluating the integral in the exponent, we thus utilize the numerical technique
developed in the context of the Boltzmann Monte Carlo method for phonon scattering events \cite{Jacoboni},
which consists of generating a random number $r$ in the way described above to find
when a (creation) event happens.

\bigskip

In practical calculations, one starts from an ensemble of particles distributed in the phase-space according to some
specified initial quasi-distribution function, which may be obtained, for example, from a particular wave-function
or density matrix describing the initial conditions of the experiment (but not restricted to these mathematical objects only).
It is trivial to generalize the operator $\hat{S}$ to act on a set of $N$ signed particles (with $N$ an arbitrary natural number)
by applying the above algorithm to every single particle in the given set.
Furthermore, if a time interval $\left[t,t+\Delta t \right]$ is provided, the operator $\hat{S}$ can be applied repeatedly to
every particle appearing in the creation process until the final time $t+\Delta t$ is reached. The final set of particles
is, therefore, constituted of every particle involved in the process described above.
It is also important to mention that practical use of the third postulate makes sense only when
the time and phase-space are discretized and two signed particles annihilate when they belong to the same cell and have opposite signs
at the same time step. Thus, for computational purposes involving the postulates II and III, discretized
time and phase-space have to be introduced (although this does not preclude the possibility of an analytical/theoretical treatment
on a {\sl{continuum}} phase-space in certain circumstances).

\bigskip

For the sake of clarity, a schematic of the algorithm is presented below (the reader is also encouraged
to download the working source code from \cite{nano-archimedes}).
\begin{verbatim}
1.    start from an initial ensemble of signed particles
2.    compute the Wigner kernel
3.    compute the gamma function
4.    for time from 0 to final time
5.        for every particle of the ensemble
6.            compute free flight
7.            (eventually) create new pairs of signed particles
8.        end for
9.        annihilation
10.   end for
\end{verbatim}
In particular, row $1$ is equivalent to assigning initial conditions to the system, row $2$ is based
on formula (\ref{wigner-kernel}) - eventually restricted to a finite and discrete domain -,
row $3$ is obtained by means of (\ref{momentum_integral}) - eventually restricted to a finite and discrete domain -,
and, finally, the nested loops starting on row $5$ are equivalent to the recursive application of the operator $\hat{S}$.
Moreover, the reader should note that row $9$ suggests the application of an annihilation step. This is
a very important point as, in certain situations, the number of signed particles may increase indefinitely if
no annihilation scheme is applied (therefore corrupting the validity of the solution).
When exactly to apply this technique during the simulation remains an open question which will be addressed elsewhere.

%

\bigskip

{\sl{On macroscopic variables}}.
The signed particle formulation makes predictions by averaging a macroscopic variable
$A=A\left( {\bf{x}}; {\bf{p}}\right)$ over an initial set of particles, taking into account their sign, i.e.
\begin{equation}
 <A> = \frac{1}{N} \sum_{i=1}^N s_i A\left( {\bf{x}}_i; {\bf{p}}_i \right),
\label{macroscopic-value}
\end{equation}
where $s_i$, ${\bf{x}}_i$ and ${\bf{p}}_i$ are the sign, the position and the momentum, respectively, of the $i$-th particle.
Some well-known examples of macroscopic variable are $A\left( {\bf{x}}; {\bf{p}}\right)=1$ which corresponds to the probability density,
$A\left( {\bf{x}}; {\bf{p}}\right)=\frac{{\bf{p}}^2}{2m}$ which corresponds to the kinetic energy, etc.
This is similar to what is done in classical statistical mechanics within the context of the
Boltzmann Monte Carlo method \cite{Jacoboni}.
Given a function $A=A\left( {\bf{x}}; {\bf{p}}\right)$, formula (\ref{macroscopic-value}) represents the {\sl{only}}
meaningful way to make predictions in this theory and which can be compared to experimental measurements.
Finally, one should note that the sum in (\ref{macroscopic-value}) runs over the whole set of $N$ particles.
Any other use involving subsets of {\sl{signed}} particles is not allowed in the theory.

\begin{figure}[h!]
\centering
\begin{minipage}{0.96\textwidth}
\begin{tabular}{c}
\includegraphics[width=0.51\textwidth]{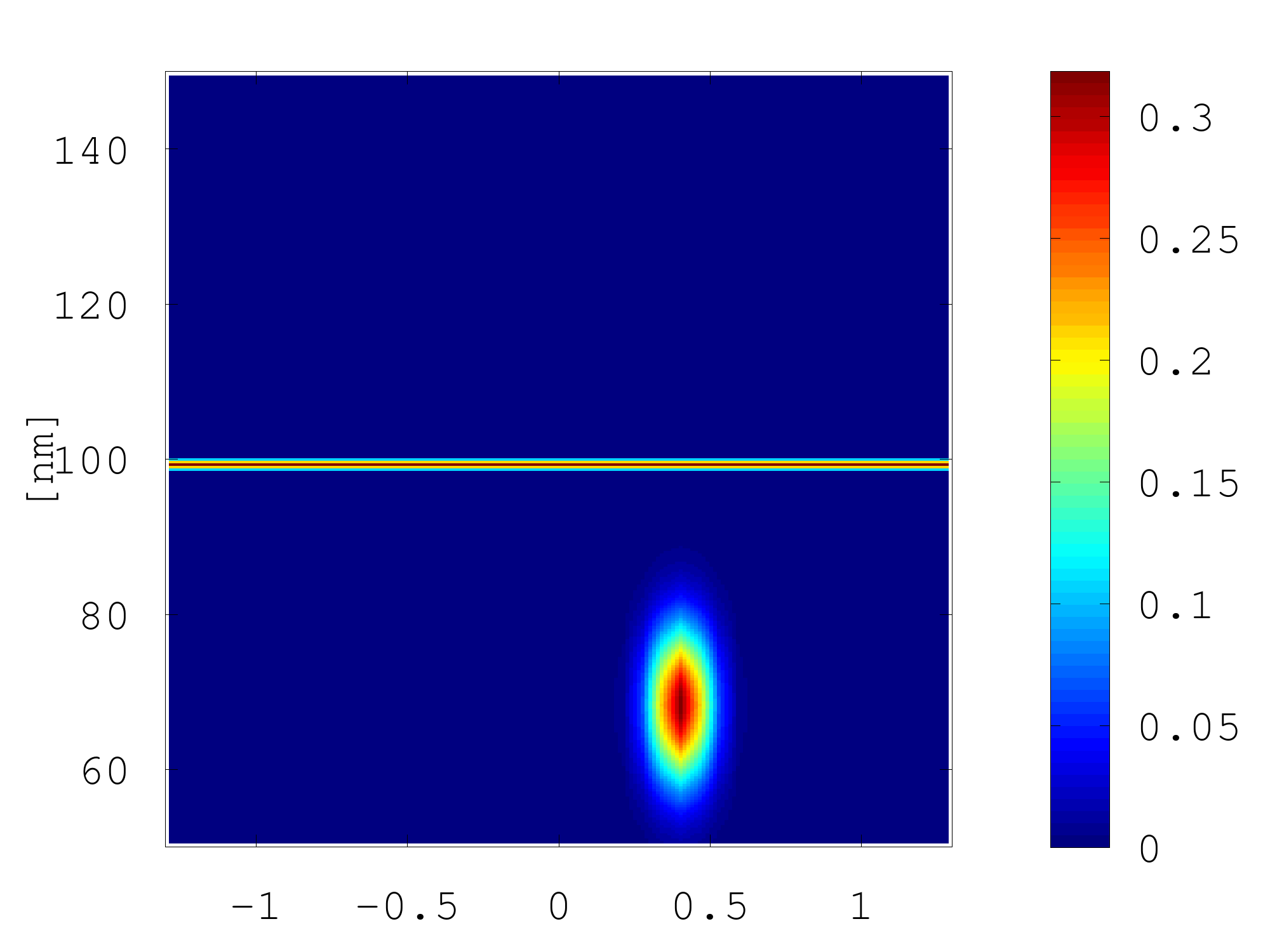}
\includegraphics[width=0.51\textwidth]{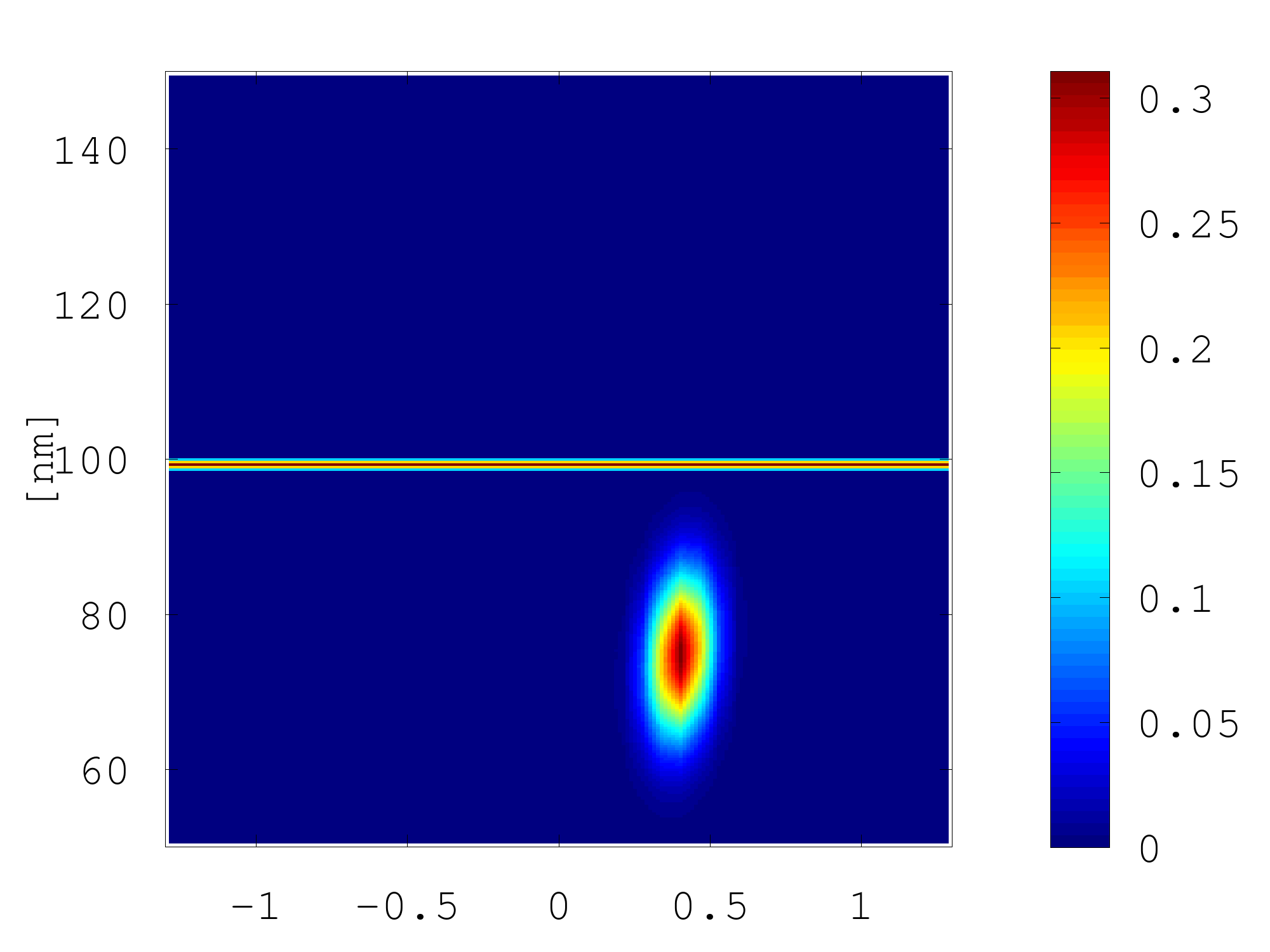}
\\
\includegraphics[width=0.51\textwidth]{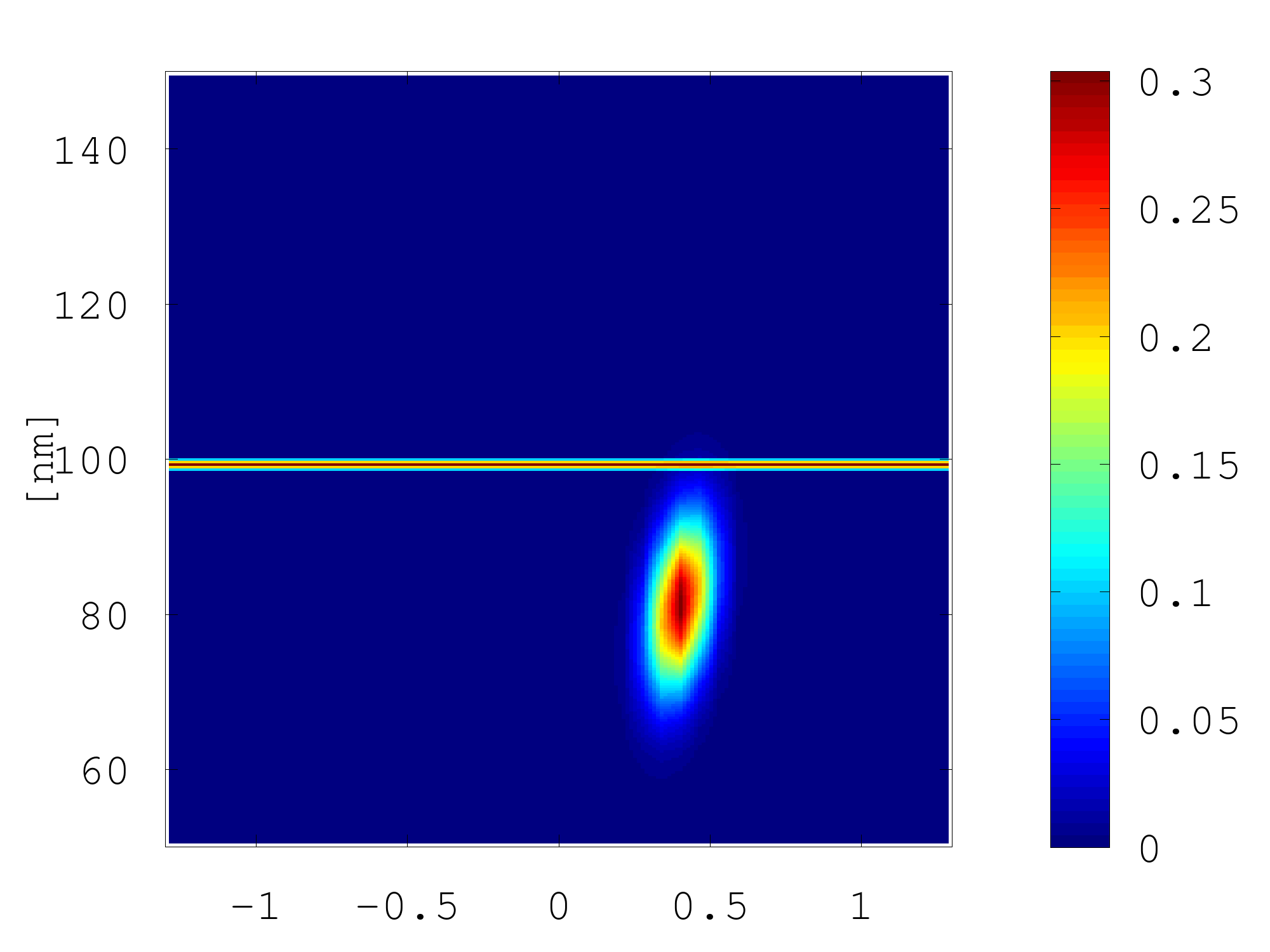}
\includegraphics[width=0.51\textwidth]{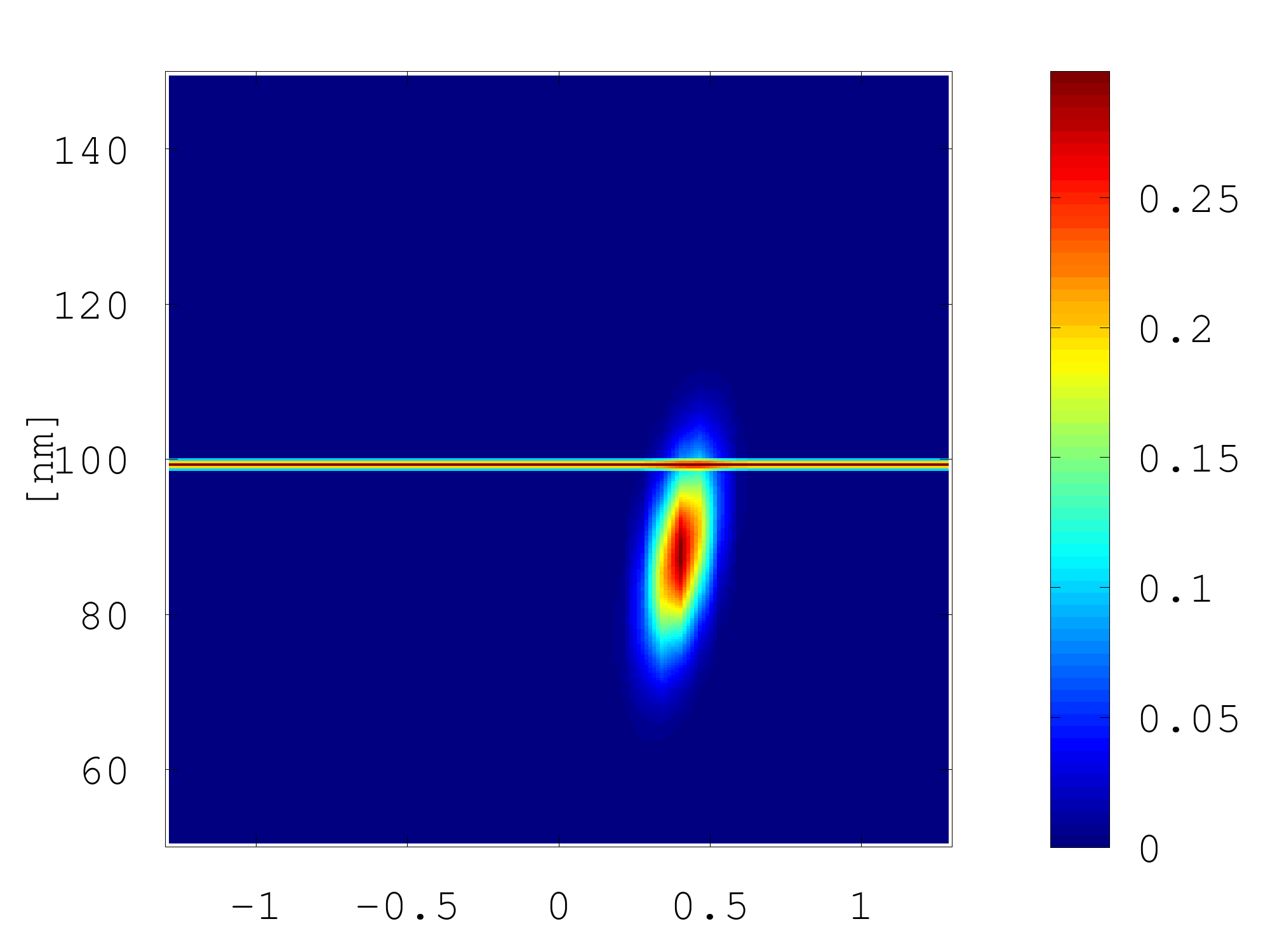}
\\
\includegraphics[width=0.51\textwidth]{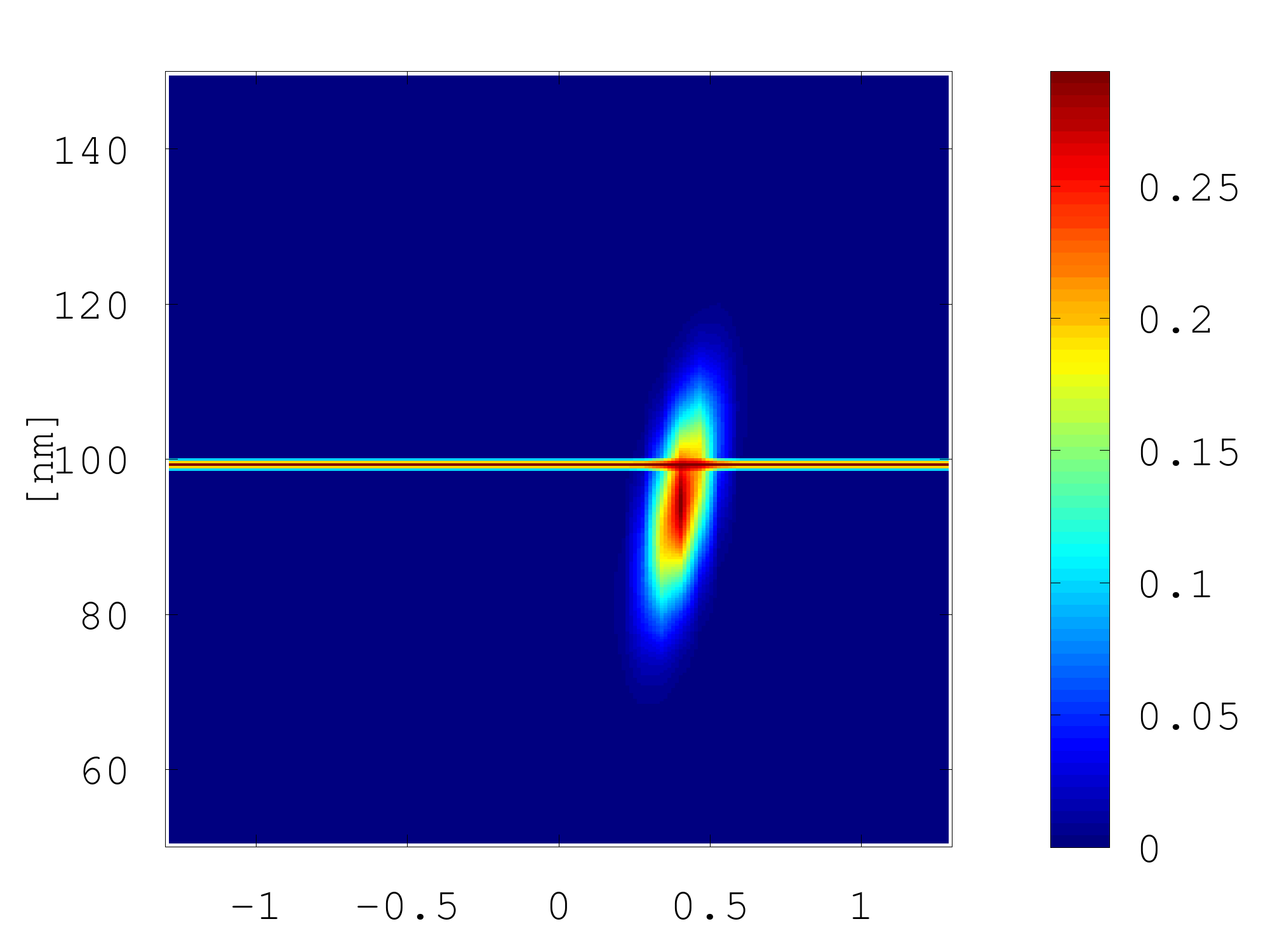}
\includegraphics[width=0.51\textwidth]{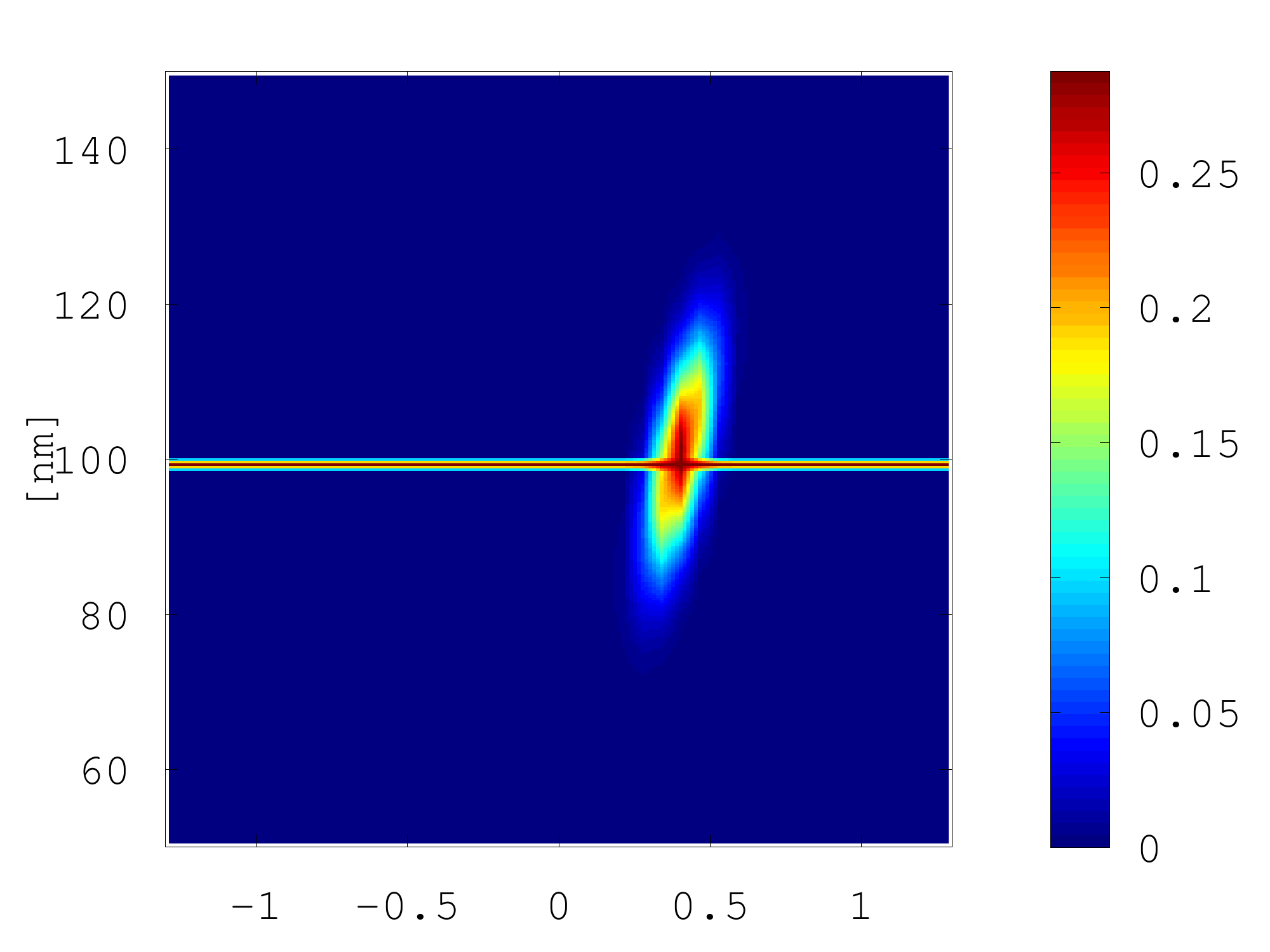}
\\
\includegraphics[width=0.51\textwidth]{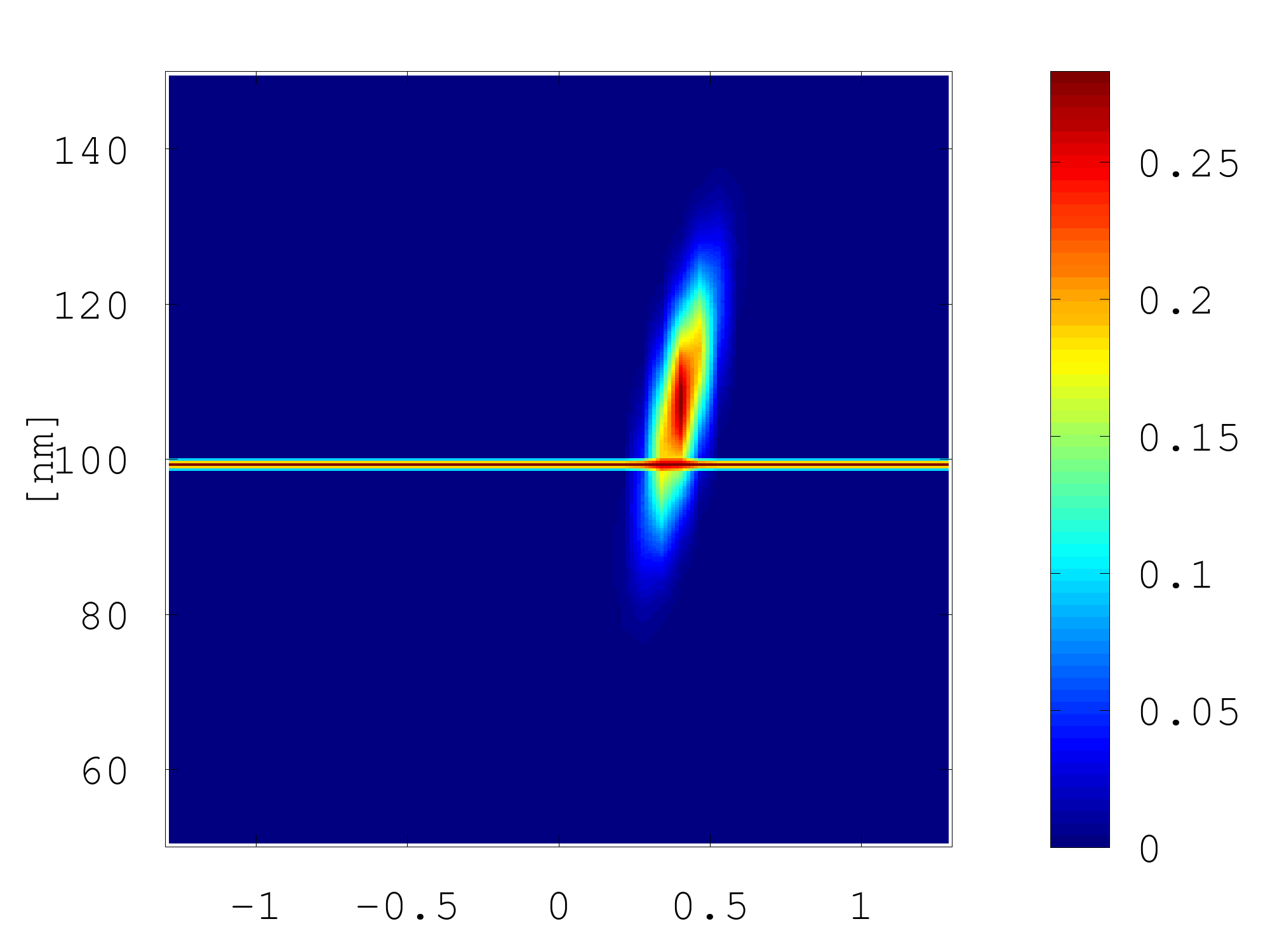}
\includegraphics[width=0.51\textwidth]{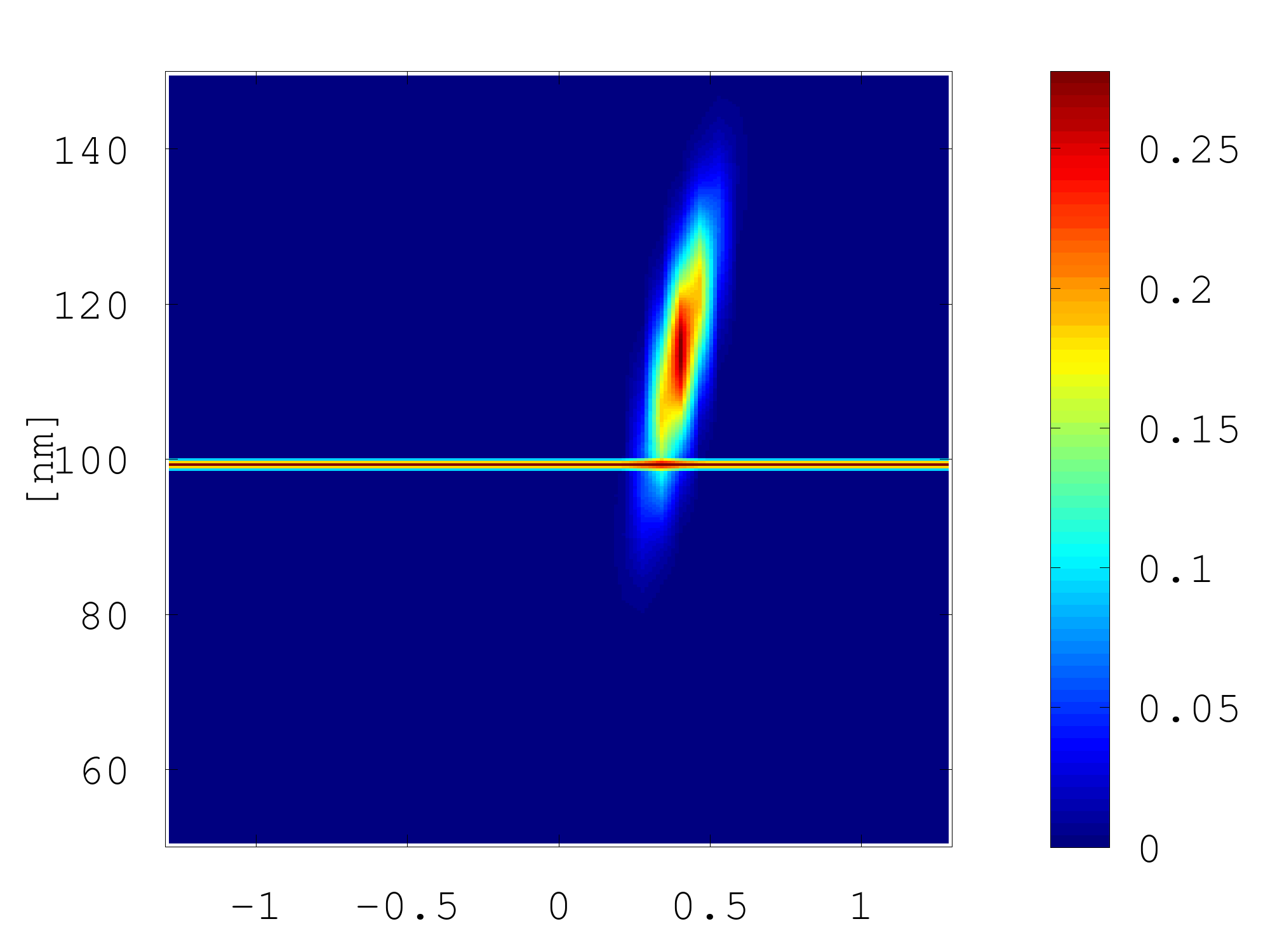}
\end{tabular}
\end{minipage}
\caption{Quantum tunnelling experiment in phase-space: an initially Gaussian wave-packet is evolving in free space. The wave packet is represented by a set of signed particles (initially all positive) which is evolved in time by repeatedly applying the operator $\hat{S}$. This plot shows the simulation at times (from left to right, from top to bottom) $0$ fs, $10$ fs, $20$ fs, $30$ fs, $40$ fs, $50$ fs, $60$ fs and $70$fs respectively. The (red) line represents the middle of the spatial domain.}
\label{distribution_00}
\end{figure}

\begin{figure}[h!]
\centering
\begin{minipage}{0.96\textwidth}
\begin{tabular}{c}
\includegraphics[width=0.51\textwidth]{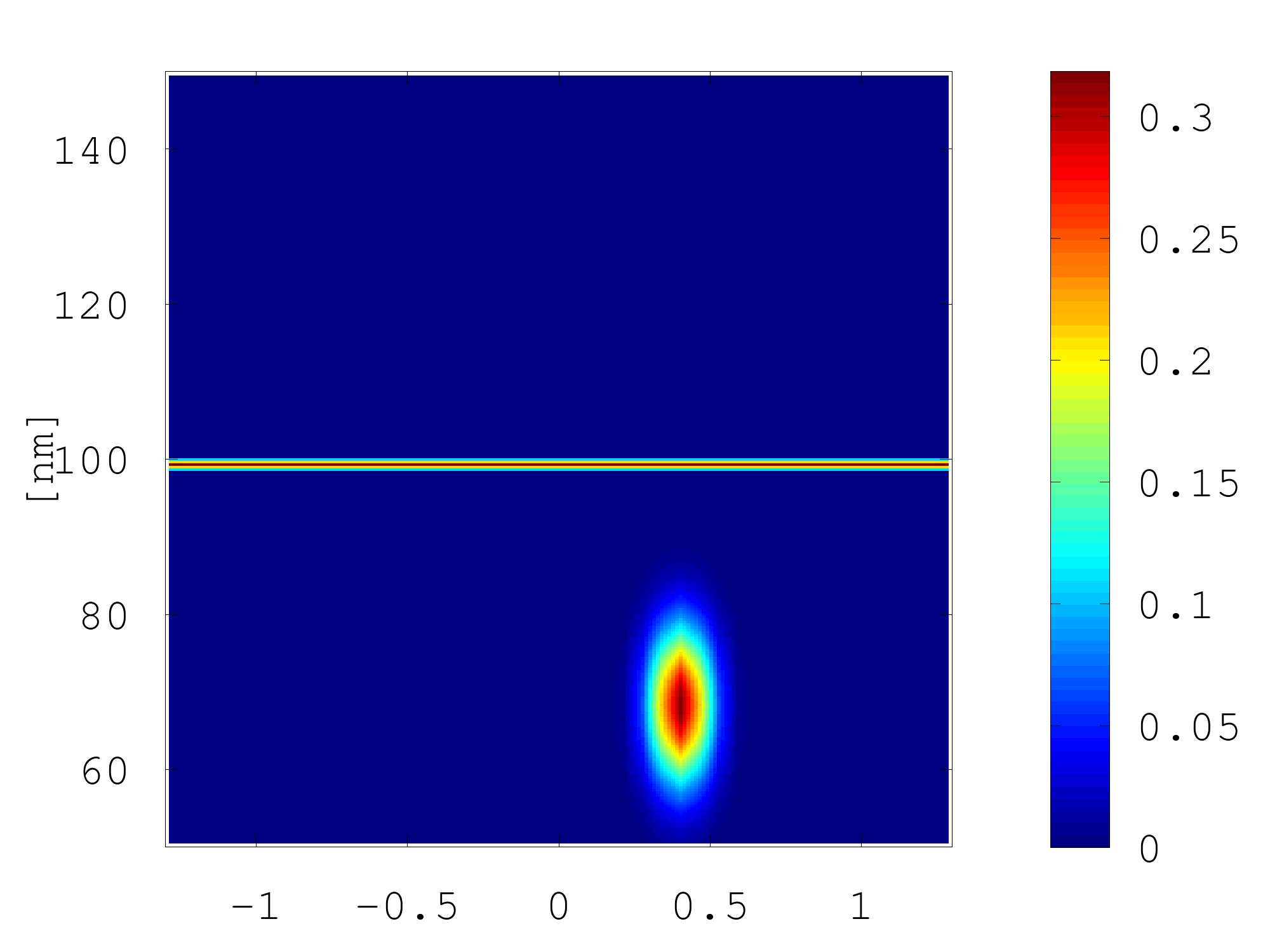}
\includegraphics[width=0.51\textwidth]{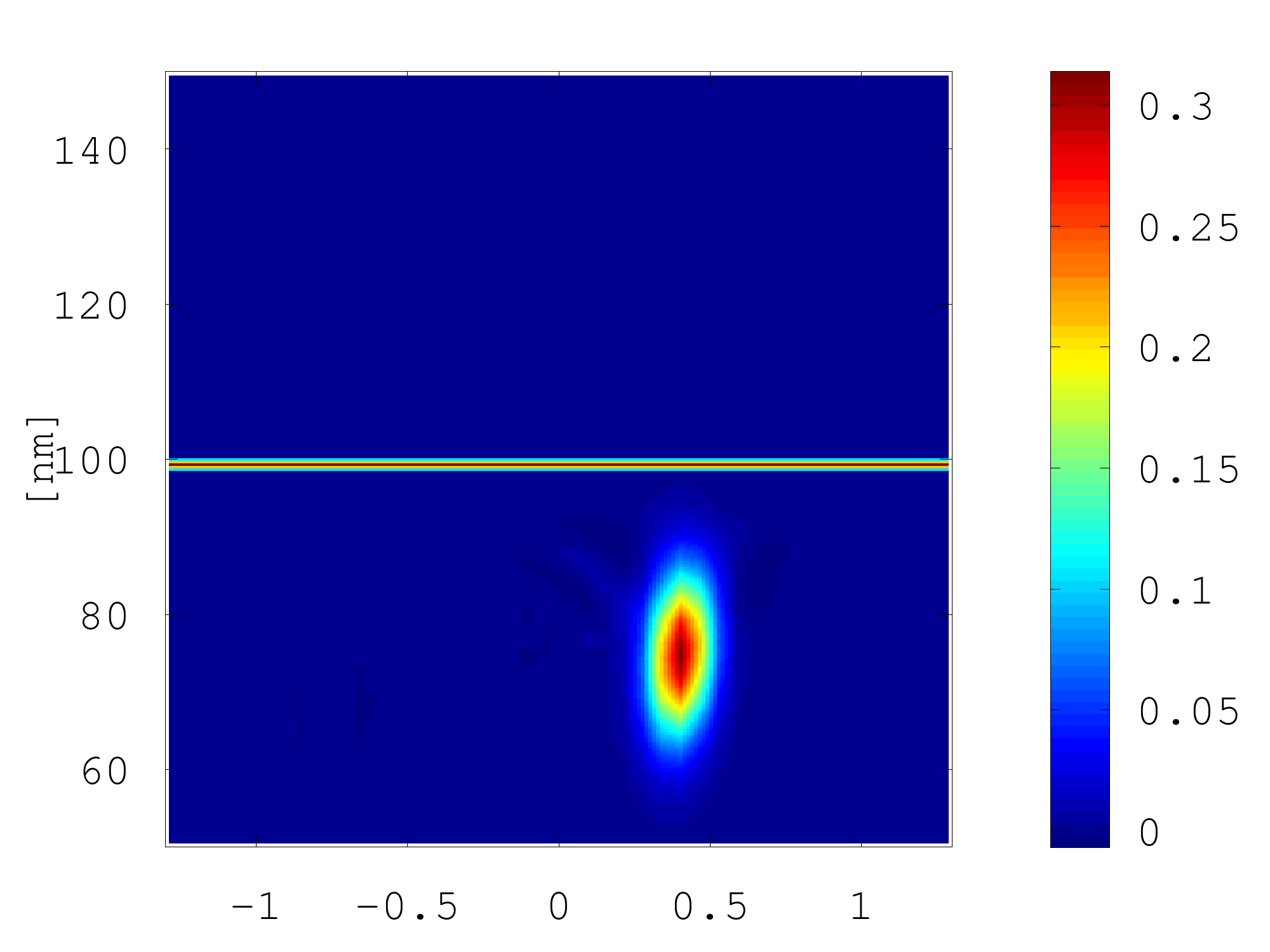}
\\
\includegraphics[width=0.51\textwidth]{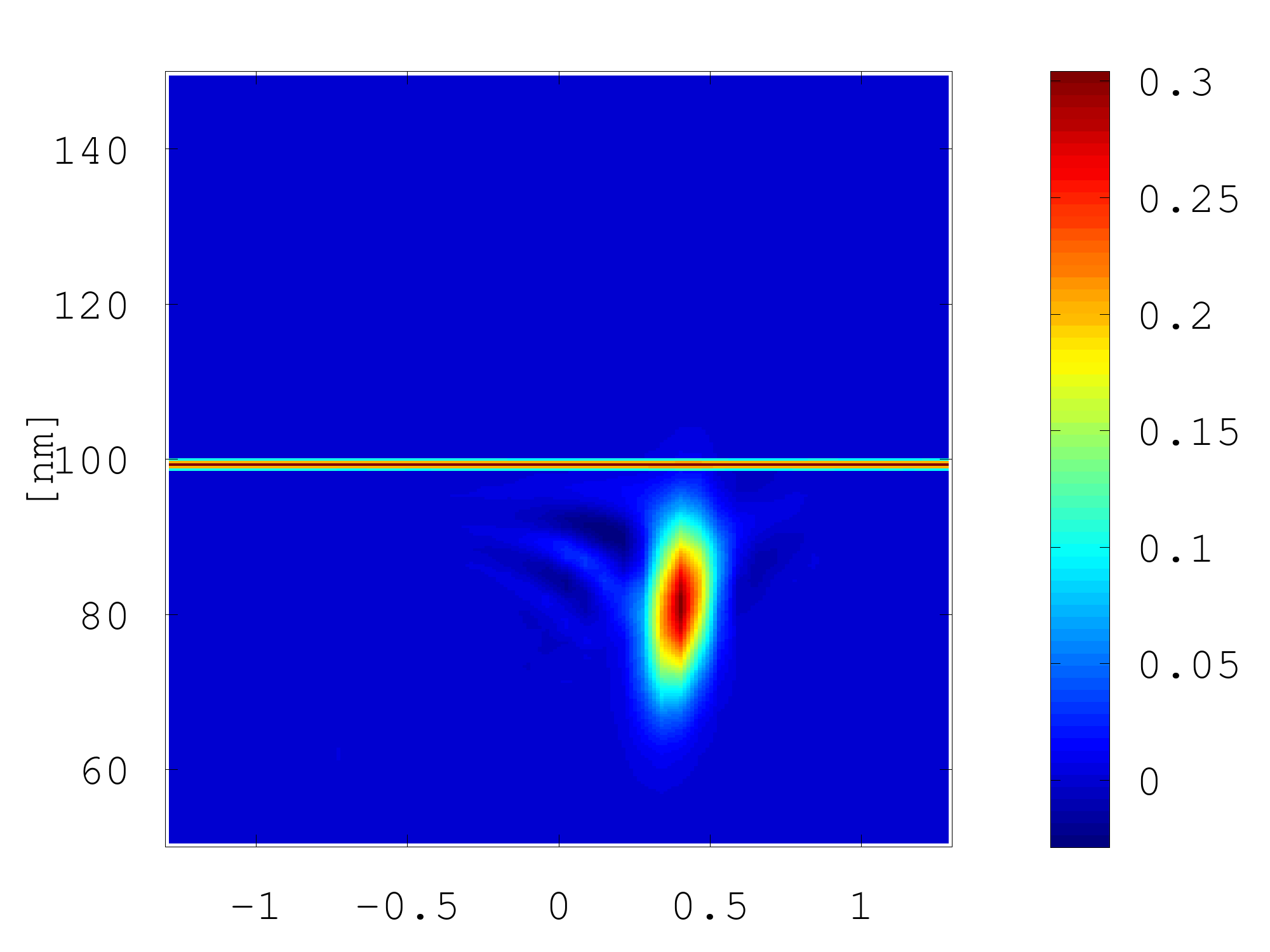}
\includegraphics[width=0.51\textwidth]{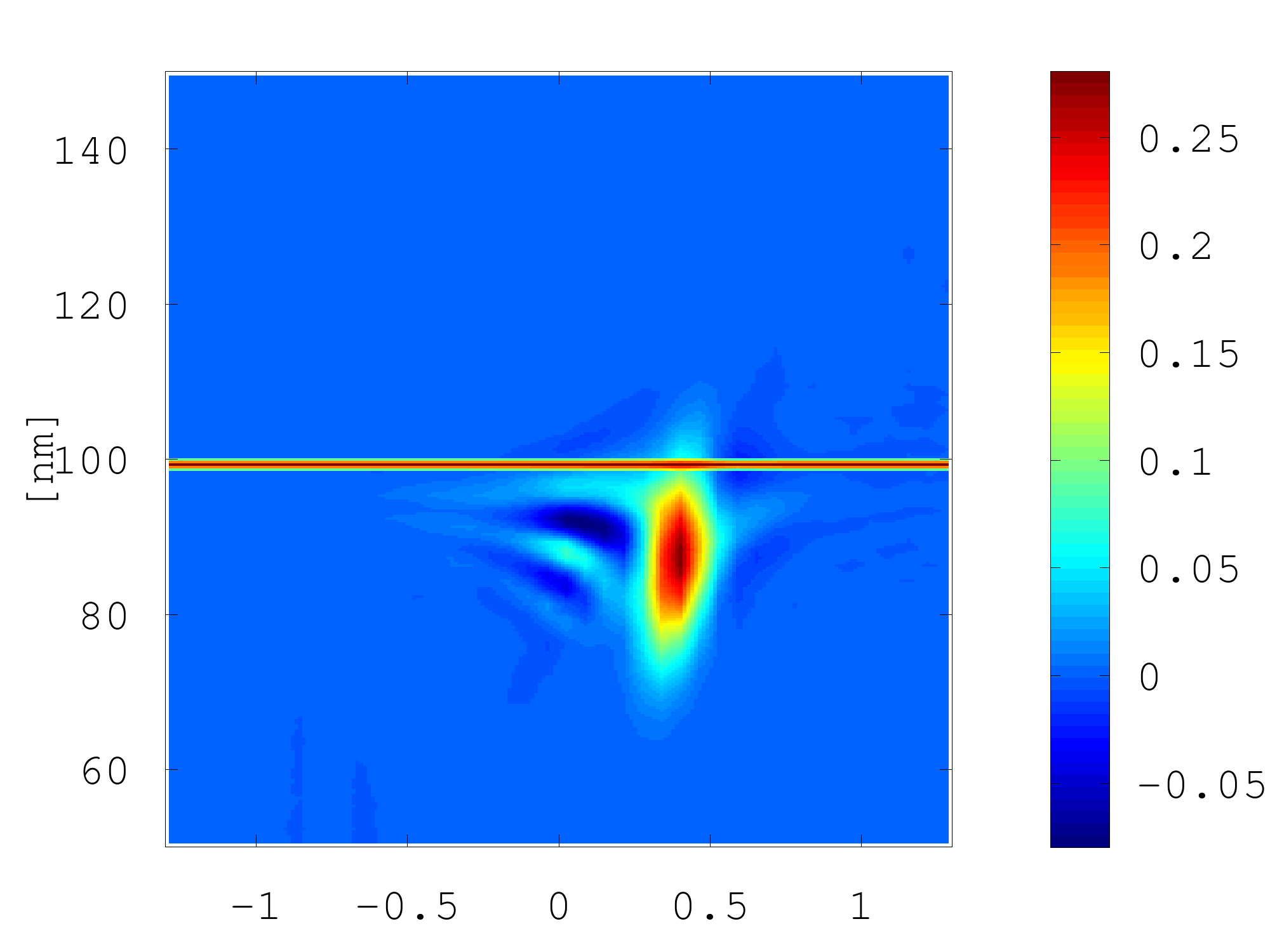}
\\
\includegraphics[width=0.51\textwidth]{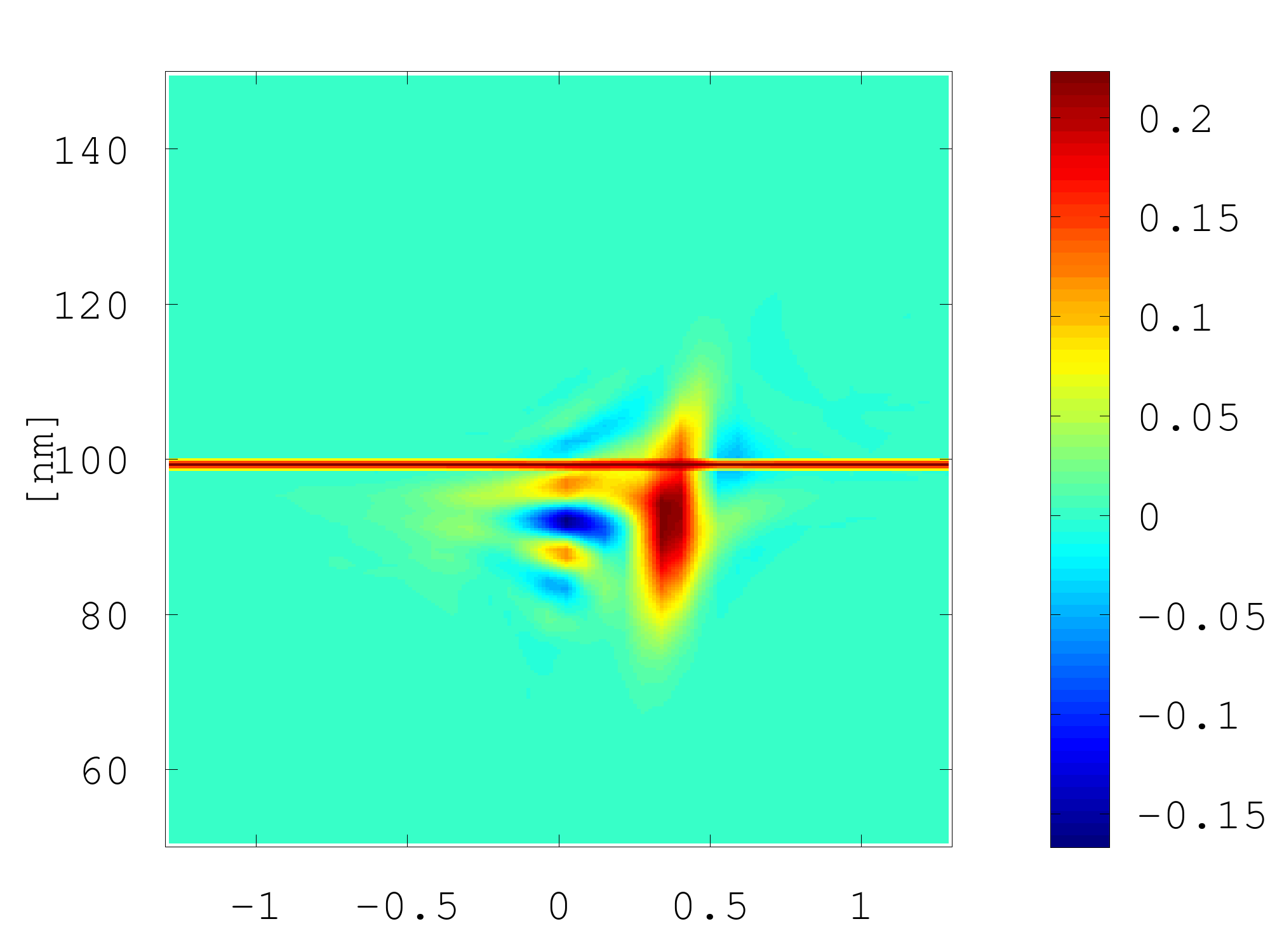}
\includegraphics[width=0.51\textwidth]{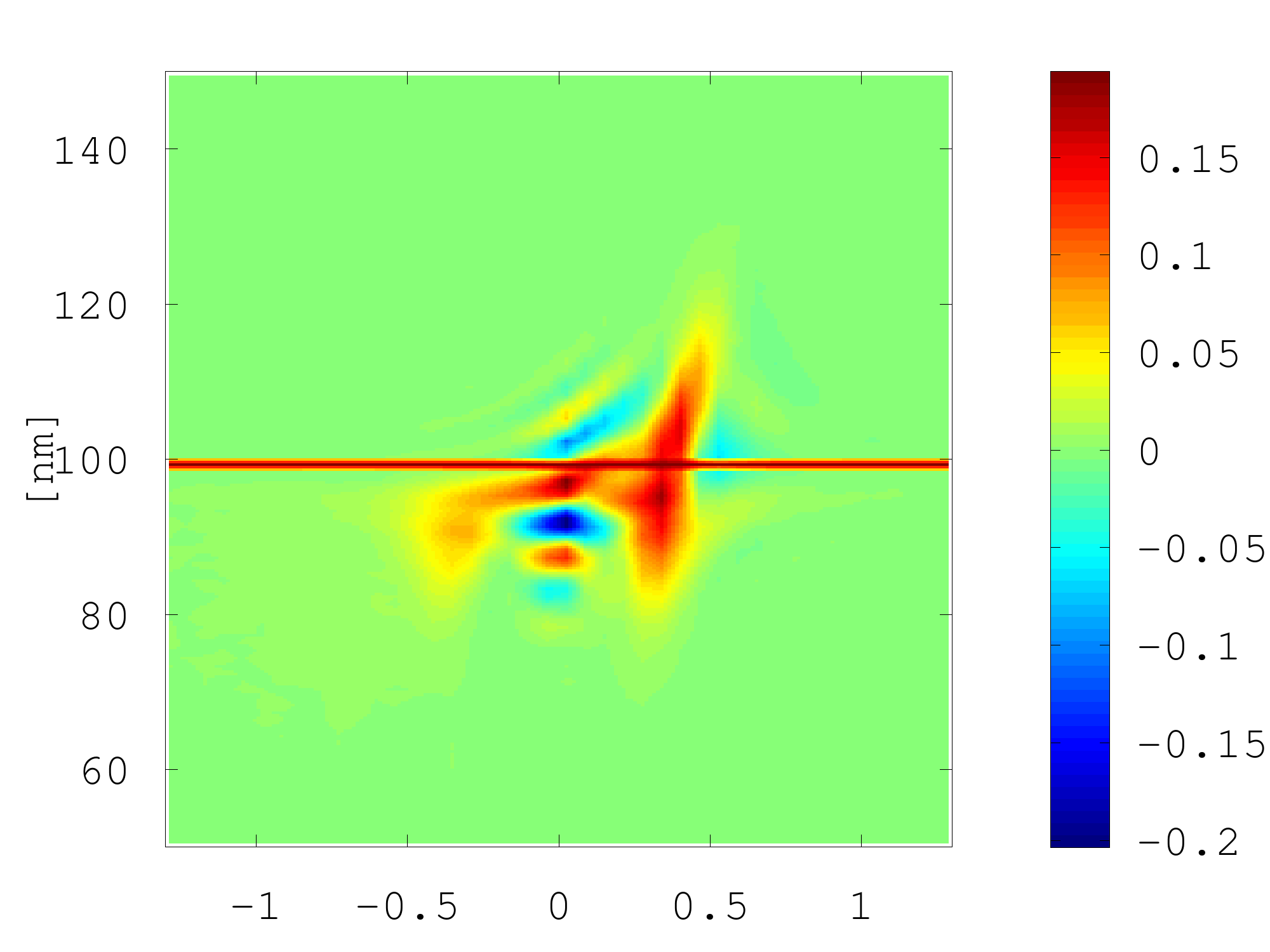}
\\
\includegraphics[width=0.51\textwidth]{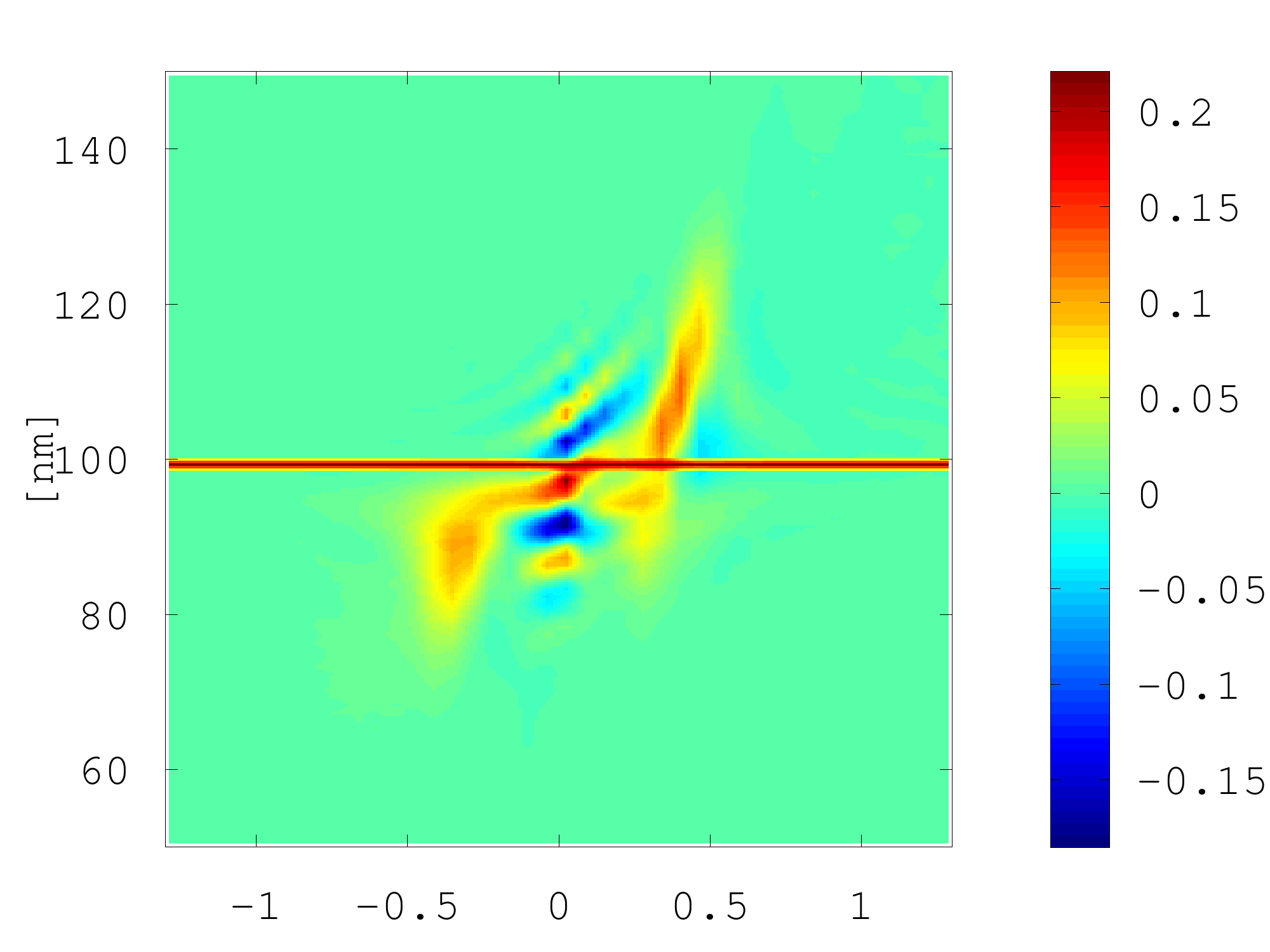}
\includegraphics[width=0.51\textwidth]{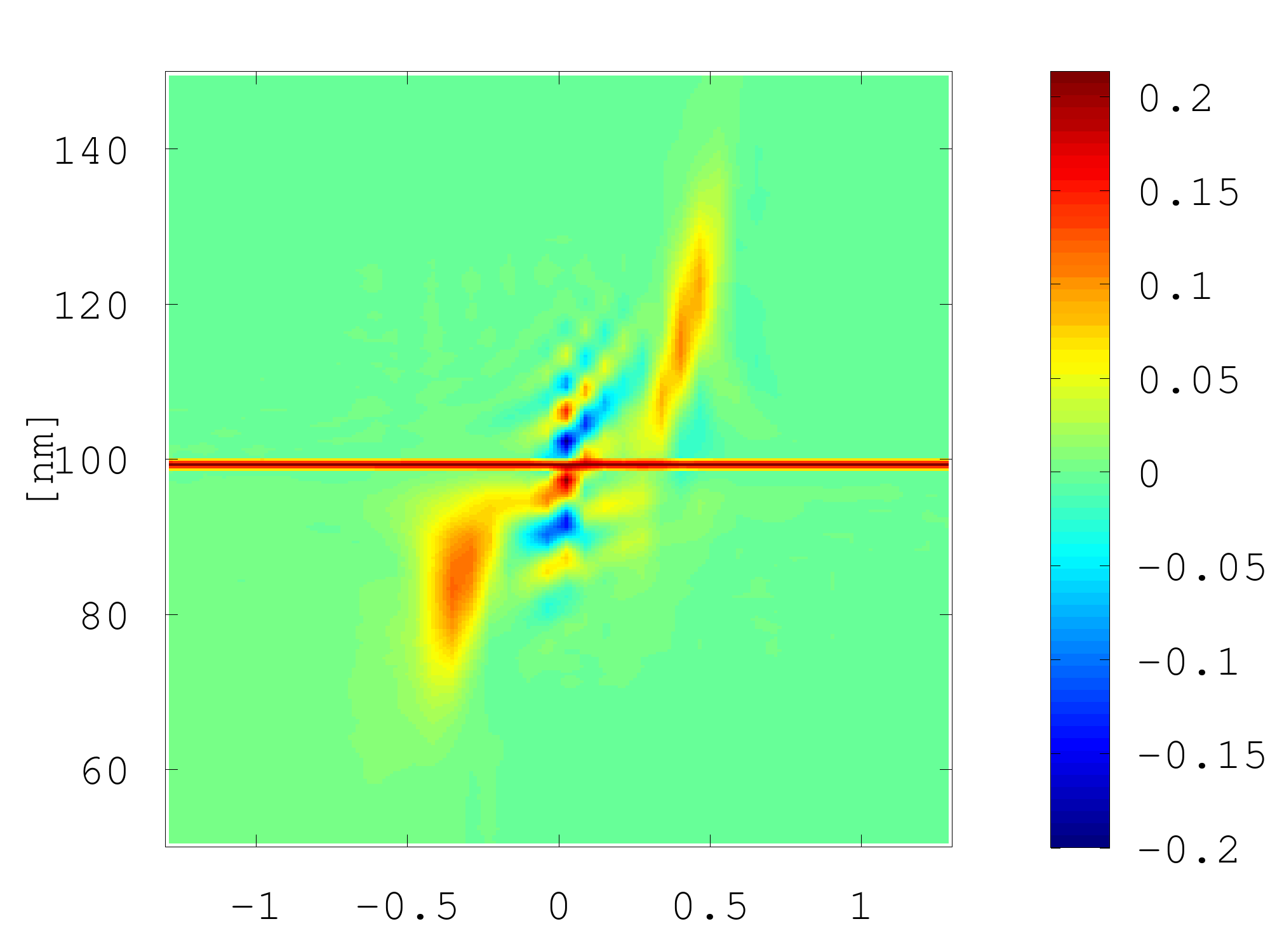}
\end{tabular}
\end{minipage}
\caption{Quantum tunnelling experiment in phase-space: an initially Gaussian wave-packet is directed towards a potential barrier equal to $0.10$ eV. The wave packet is represented by a set of signed particles (initially all positive) which is evolved in time by repeatedly applying the operator $\hat{S}$. This plot shows the simulation at times (from left to right, from top to bottom) $0$ fs, $10$ fs, $20$ fs, $30$ fs, $40$ fs, $50$ fs, $60$ fs and $70$fs respectively. One notes the appearance of negative values (dark blue) during the evolution. The position of the potential barrier is schematically represented by the (red) line in the middle of the domain.}
\label{distribution_01}
\end{figure}

\begin{figure}[h!]
\centering
\begin{minipage}{0.96\textwidth}
\begin{tabular}{c}
\includegraphics[width=0.51\textwidth]{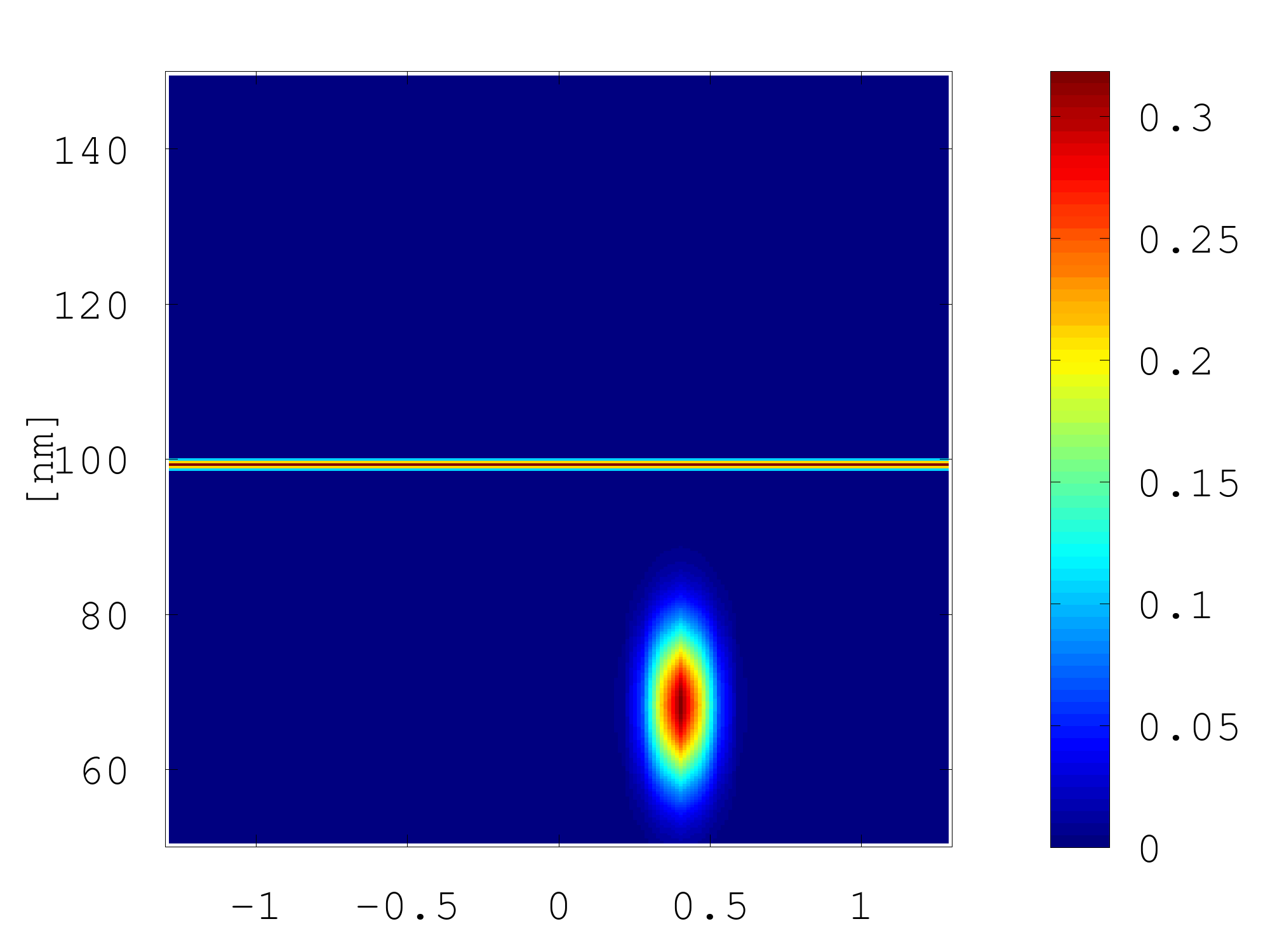}
\includegraphics[width=0.51\textwidth]{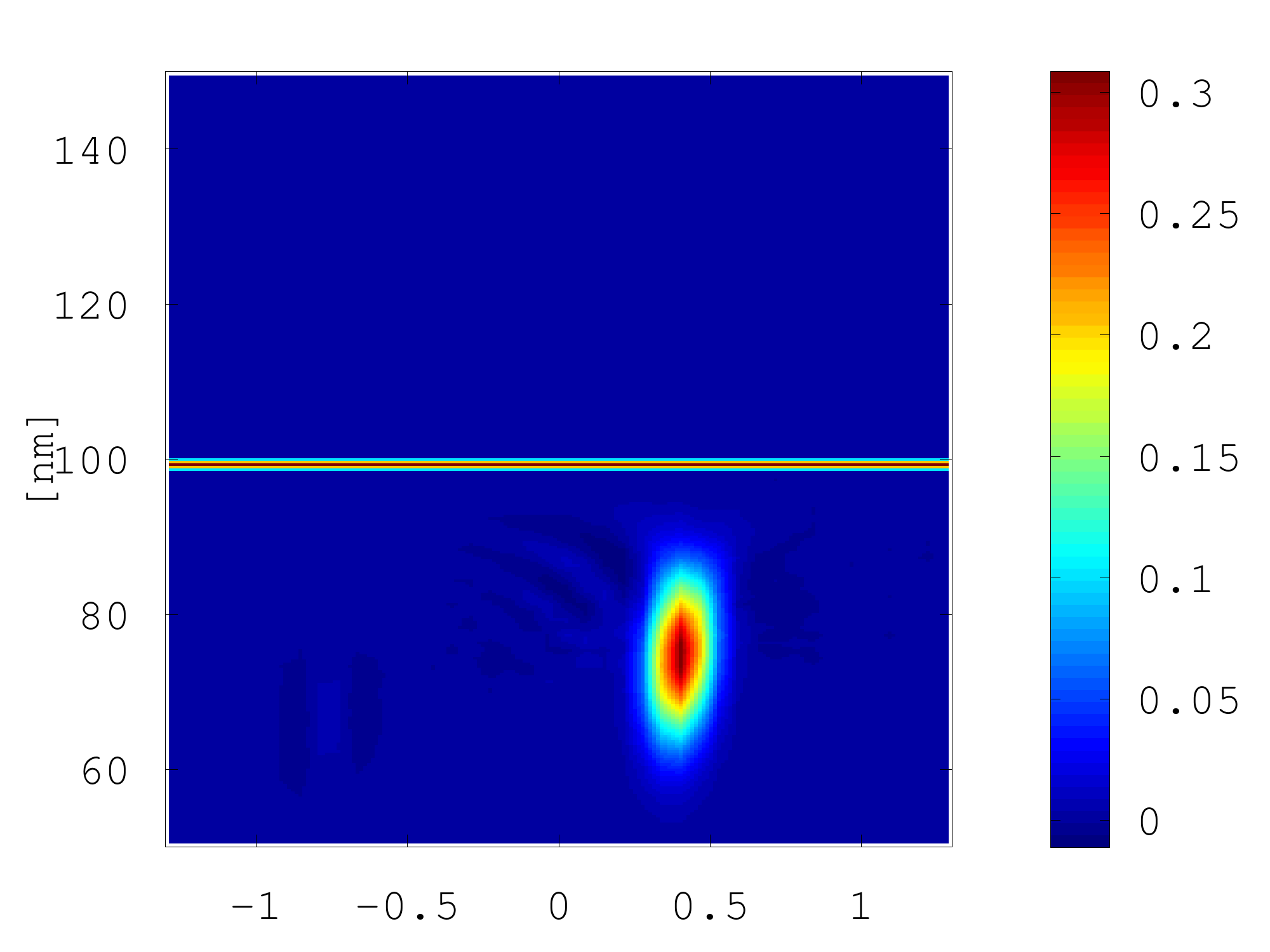}
\\
\includegraphics[width=0.51\textwidth]{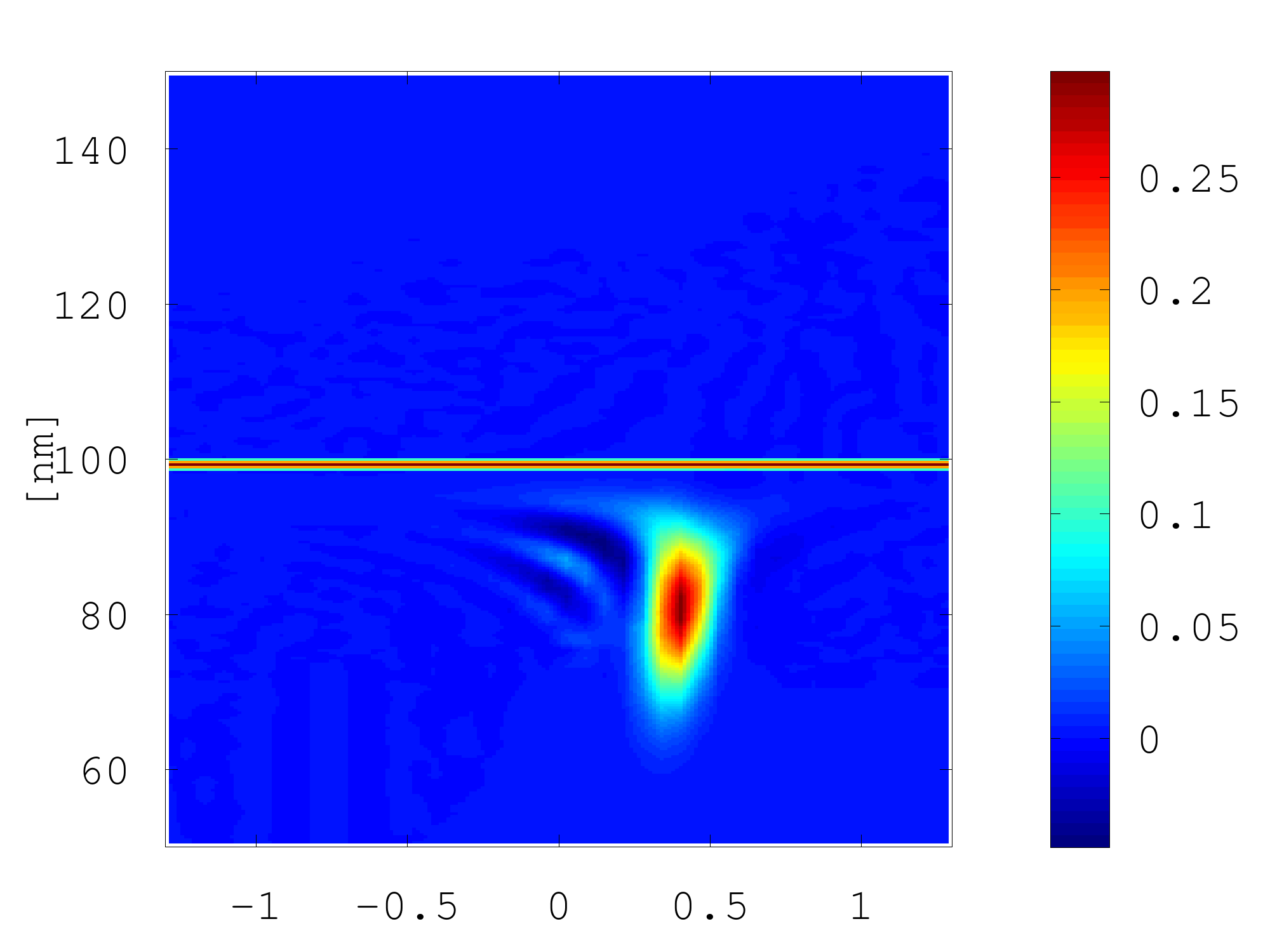}
\includegraphics[width=0.51\textwidth]{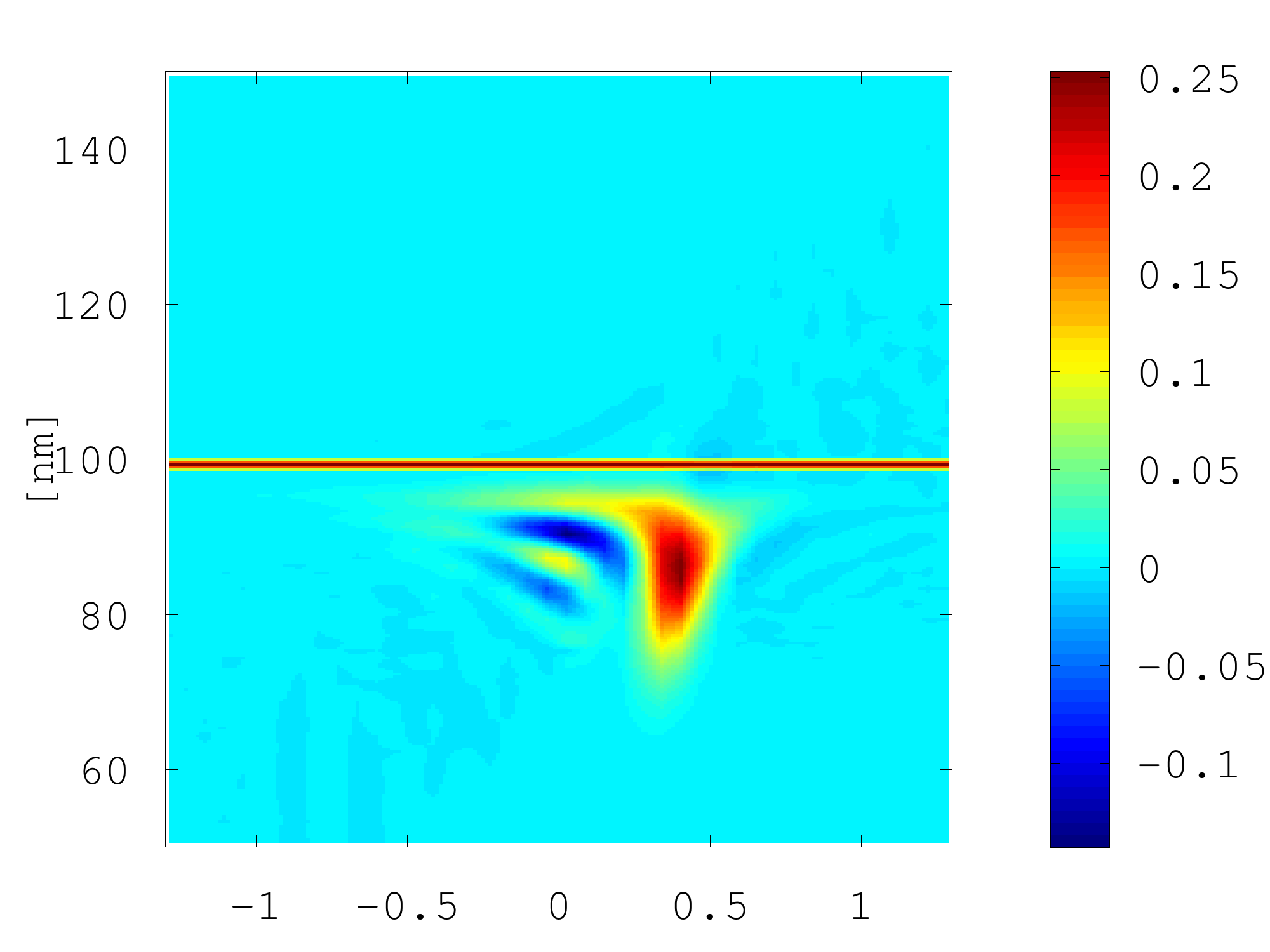}
\\
\includegraphics[width=0.51\textwidth]{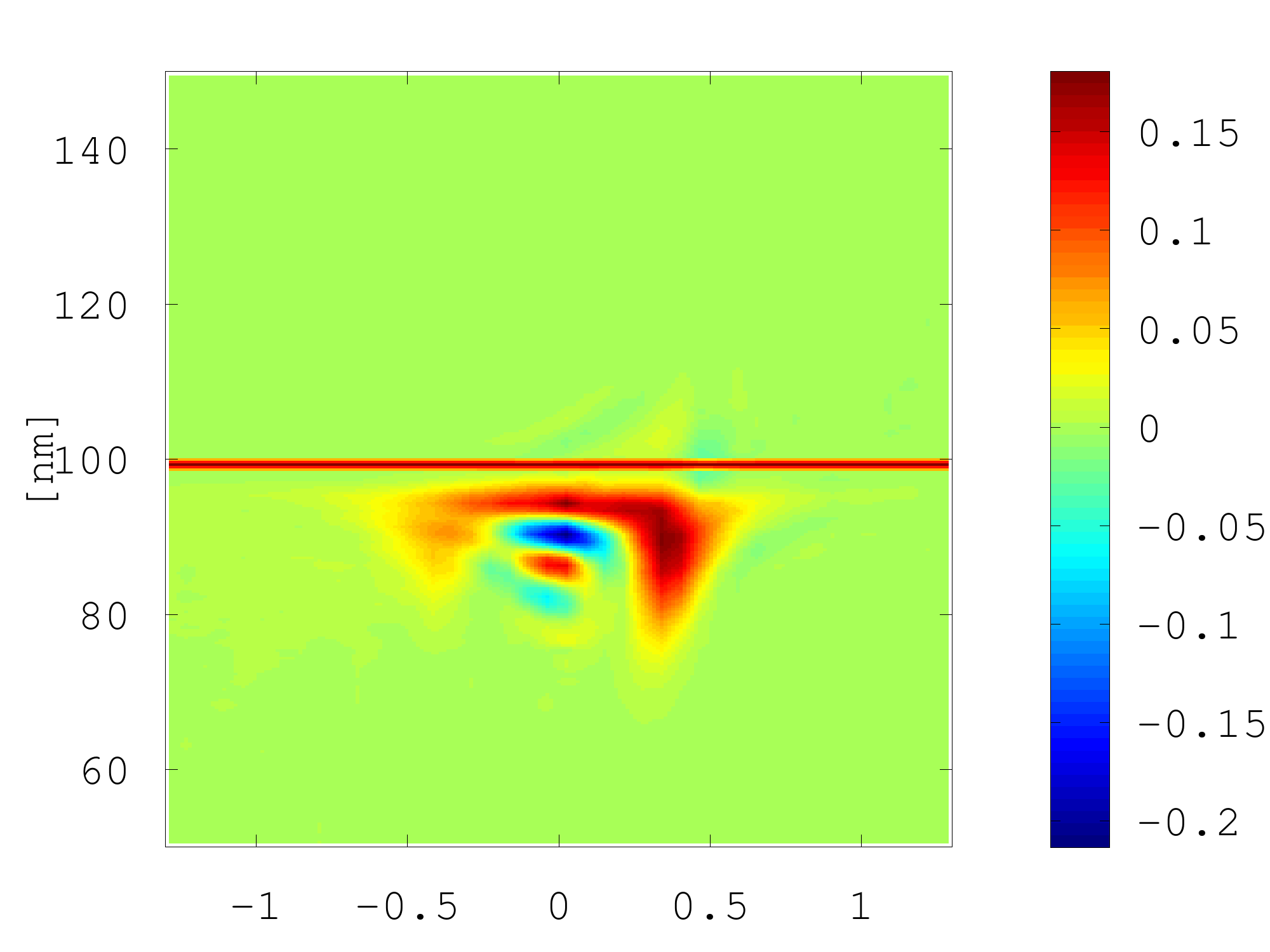}
\includegraphics[width=0.51\textwidth]{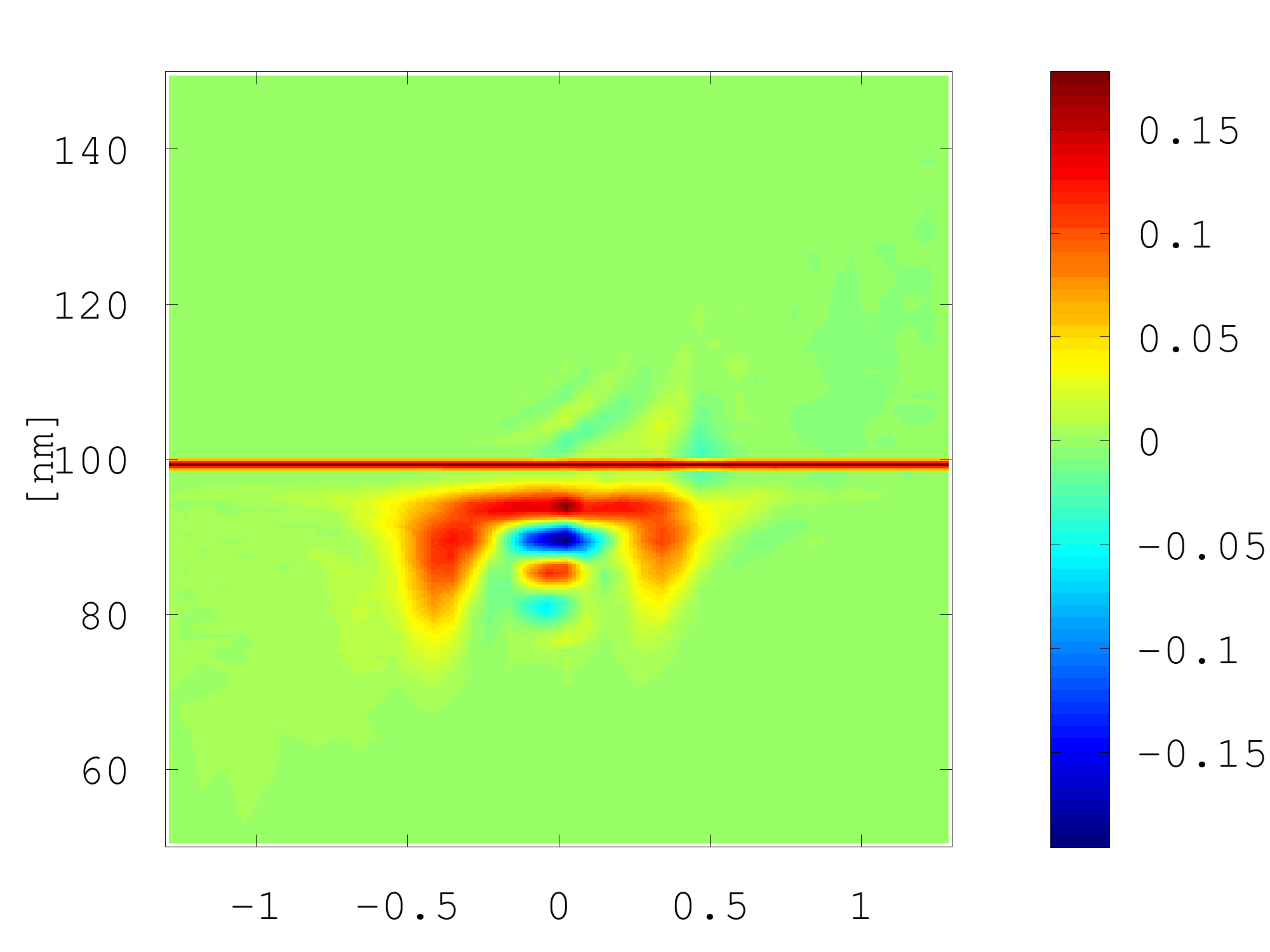}
\\
\includegraphics[width=0.51\textwidth]{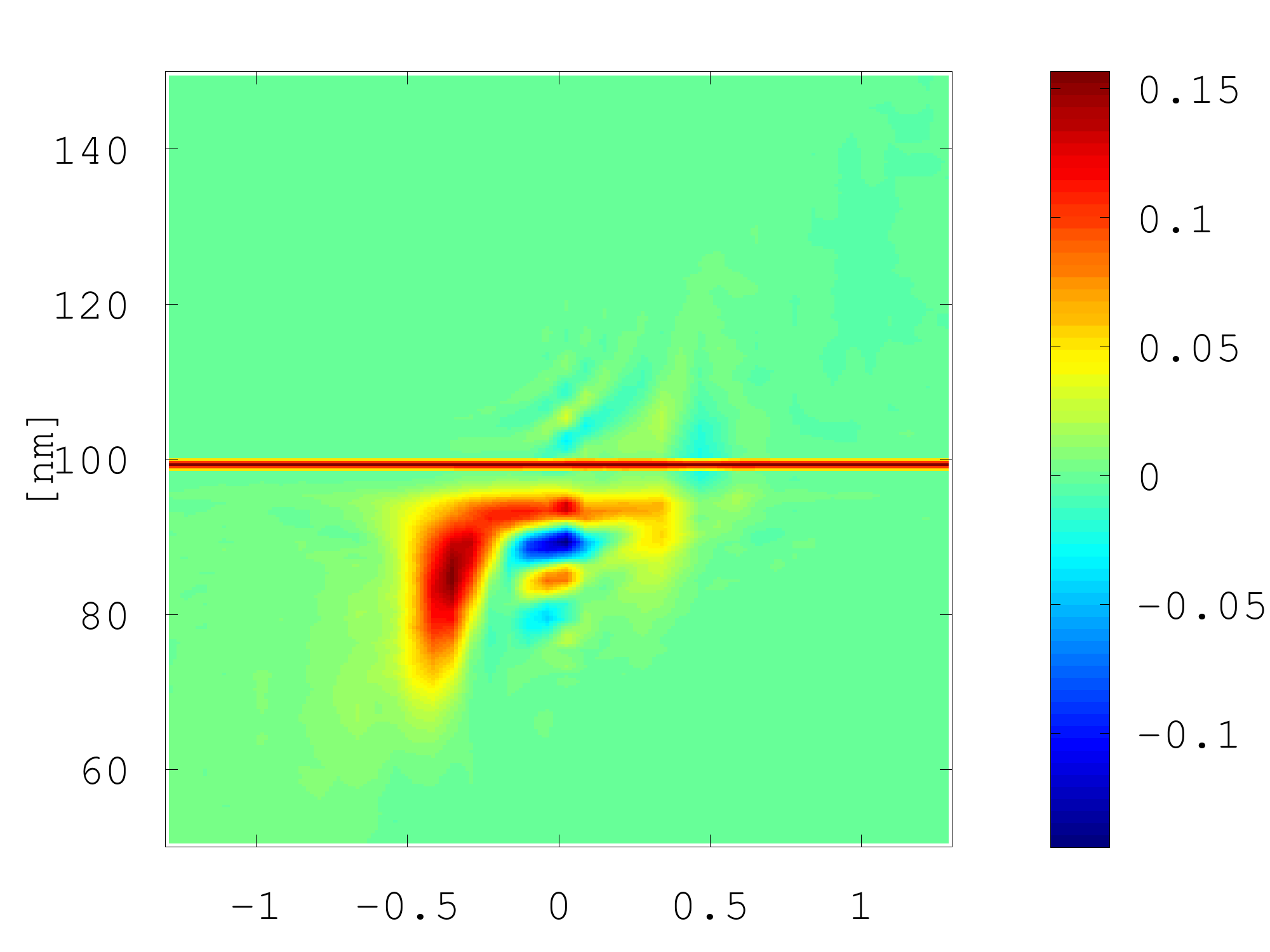}
\includegraphics[width=0.51\textwidth]{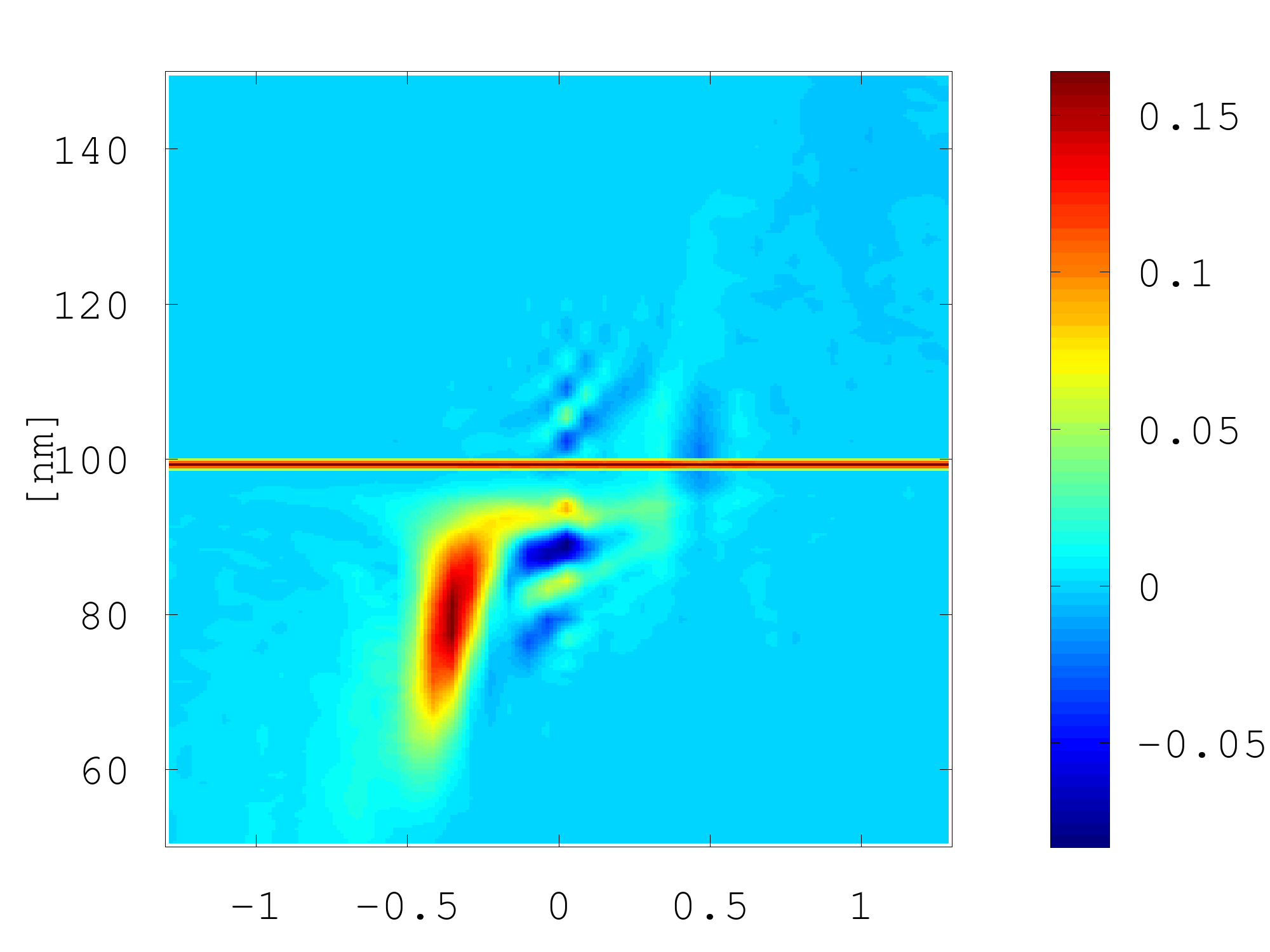}
\end{tabular}
\end{minipage}
\caption{Quantum tunnelling experiment in phase-space: an initially Gaussian wave-packet is directed towards a potential barrier equal to $0.30$ eV. The wave packet is represented by a set of signed particles (initially all positive) which is evolved in time by repeatedly applying the operator $\hat{S}$. This plot shows the simulation at times (from left to right, from top to bottom) $0$ fs, $10$ fs, $20$ fs, $30$ fs, $40$ fs, $50$ fs, $60$ fs and $70$fs respectively. One notes the appearance of negative values (dark blue) during the evolution. The position of the potential barrier is schematically represented by the (red) line in the middle of the domain.}
\label{distribution_03}
\end{figure}

\section{Derivation of the postulates}

In this section, we show that the set of three postulates defined in the previous section
has its origins in the time-dependent single-body Wigner equation \cite{Wigner}.
In fact, algebraic manipulations of this equation endorse a unique physical interpretation
which, eventually, offers a novel physical picture on the dynamics of quantum systems.
In a sense, this is not any different from other formulations which come from previous formulations
(e.g. the Wigner formalism is directly obtained from the time-dependent Schr\"{o}dinger equation,
although it allows a totally different perspective).
Finally we show that, in this context, the signed particle Wigner Monte Carlo (MC) method
(defined and thoroughly described in \cite{MCMA}, \cite{JCP-01}, \cite{JCP-02})
can be naturally interpreted as a numerical discretization of the new formulation (by introducing a semi-discrete phase-space),
restricted to a finite domain.

\bigskip

The time-dependent single-body Wigner equation \cite{Wigner} describes
the evolution of a quasi-distribution function $f_W=f_W({\bf{x}}; {\bf{p}}; t)$
defined over a phase-space and reads:
\begin{equation}
 \frac{\partial f_W}{\partial t}({\bf{x}}; {\bf{p}}; t) + \frac{{\bf{p}}}{m} \cdot \nabla_{{\bf{x}}} f_W
= \int d{\bf{p}}' f_W({\bf{x}}; {\bf{p}} + {\bf{p}}'; t) V_W({\bf{x}}; {\bf{p}}'; t),
\label{Wigner_equ}
\end{equation}
with $V_W=V_W({\bf{x}}; {\bf{p}}; t)$ the Wigner kernel (\ref{wigner-kernel}) \cite{Wigner}.
It is possible to show that the prediction made by means of this equation are exactly the same as
the ones provided by the Schr\"{o}dinger equation (through the invertible Wigner-Weyl transform) \cite{Dias}.
Therefore, solving the Wigner equation is equivalent to solve the Schr\"{o}dinger equation, despite
they provide two very different pictures.


Through simple algebraic manipulations (by adding to both sides of the equation the quantity $\gamma({\bf{x}}(y))$ and reordering the terms),
it is possible to express the Wigner equation (\ref{Wigner_equ}) in a integral form,
more precisely as a Fredholm equation of second kind:
\begin{eqnarray}
&&f_W({\bf{x}}; {\bf{p}}; t)-
e^{-\int_{0}^t\gamma({\bf{x}}(y))dy}
f_i({\bf{x}}(0); {\bf{p}})
=
\label{backwardforma}
\\
&&
\int_{0}^\infty dt'
\int_{-\infty}^{+\infty} d{\bf{p}}'
\int d{\bf{x}}'
f_W({\bf{x}}'; {\bf{p}}'; t')
\Gamma({\bf{x}}'; {\bf{p}}; {\bf{p}}')
e^{-\int_{t'}^t\gamma({\bf{x}}(y))dy}
\theta(t-t')
\delta({\bf{x}}'-{\bf{x}}(t'))\theta_D({\bf{x}}'),
\nonumber
\end{eqnarray}
where
$$
\Gamma({\bf{x}},{\bf{p}},{\bf{p}}') = V_w^+({\bf{x}},{\bf{p}}-{\bf{p}}') - V_w^+({\bf{x}},-({\bf{p}}-{\bf{p}}')) 
+ \gamma({\bf{x}}) \delta({\bf{p}}-{\bf{p}}'),
$$
with $f_i({\bf{x}}; {\bf{p}})$ formally representing the initial conditions
of the system at time $t=0$. The reader should note that no discretization of the phase-space is introduced in the process.

Now, equation (\ref{backwardforma}) suggests that it is possible to express the expectation value $\langle A\rangle$ of a macroscopic variable
$A=A \left( {\bf{x}}; {\bf{p}} \right)$ as a Liouville-Neumann series.
For instance, the zeroth order term of the series reads:
$$
\langle A\rangle_0(\tau) =
\int_0^\infty dt'
\int d{\bf{x}}
\int_{-\infty}^{+\infty} d{\bf{p}}'
f_i({\bf{x}}; {\bf{p}}' )
e^{-\int_{0}^{t'}\gamma({\bf{x}}(y))dy}
A({\bf{x}}(t'),{\bf{p}}')\delta(t'-\tau).
$$
By simple mathematical considerations, it is possible to interpret part
of the integrand as a product of conditional probabilities.
Thus, if $f_i=f_i({\bf{x}}; {\bf{p}})$ is normalized to unity, one
generates random points $({\bf{x}}; {\bf{p}}')$ in the (continuum) phase-space at the initial time $t=0$.
This process initializes the particle trajectories ${\bf{x}} = {\bf{x}}(t)$
and the exponent gives the probability for the particle to remain over the trajectory provided that
the out-of-trajectory event rate is provided by the function $\gamma = \gamma({\bf{x}})$.
This probability filters out these particles, so that the randomly generated out-of-trajectory time
is less than $\tau$. If the particle remains in the same trajectory till time $\tau$,
it has a contribution to $\langle A\rangle_0(\tau)$ equal to  $f_i({\bf{x}},{\bf{p}}' )A({\bf{x}}(\tau),{\bf{p}}')$,
otherwise it does not contribute at all.
Therefore, $\langle A\rangle_0(\tau)$ is estimated by the mean value obtained from the $N$ initialized particles.

In the same way, the first three terms of the Liouville-Neumann series show how to continue with higher order terms in a way similar to \cite{MCMA}
(but over a continuum phase-space).
Indeed, one observes that the expansion of $\langle A\rangle$ actually branches,
and the total value is given by the sum of all branches.
Considering the presence of the term $\Gamma = \Gamma({\bf{x}},{\bf{p}},{\bf{p}}')$
in the series, one can equivalently talk in terms of three appearing particles.
The initial parent particle survives and a couple of new signed particles, one positive and one negative, are generated with the first two
probabilities. This introduces the first postulate of our suggested formulation (signed particles)
and part of the second postulate (particles generation).

In the creation process, one momentum state ${\bf{p}}-{\bf{p}}'={\bf{q}}$ is generated with probability:
$$
\frac{V_W^+({\bf{x}},{\bf{q}})}{\gamma({\bf{x}})},
$$
and, with the same probability, we generate another value, say ${\bf{q}}'$,
for the second momentum state ${\bf{p}}'-{\bf{p}}={\bf{q}}'$.
Thus, after any free flight the initial particle creates
two new particles with opposite signs and momentum offset (around the initial momentum)
equal to $+{\bf{q}}$ and $-{\bf{q}}$ with ${\bf{q}}={\bf{p}}-{\bf{p}}'$.
This completes the formulation of the second postulate.
The third and last postulate (particles annihilation) is a simple consequence of the way one computes macroscopic variables, as depicted in formula (\ref{macroscopic-value}).

This concludes the derivation of the set of postulates presented in the previous section.
In this context, it now becomes clear that, if one restricts the spatial domain over a finite length
and introduce a discretization of the momentum space, the Wigner MC method is re-obtained \cite{MCMA}, \cite{JCP-02}.
Therefore, the signed particle formulation of quantum mechanics represents a generalization of this method
which can be considered a valid new physical picture of the quantum world.
%
%

\begin{figure}[h!]
\centering
\begin{minipage}{0.96\textwidth}
\begin{tabular}{c}
\includegraphics[width=1.0\textwidth]{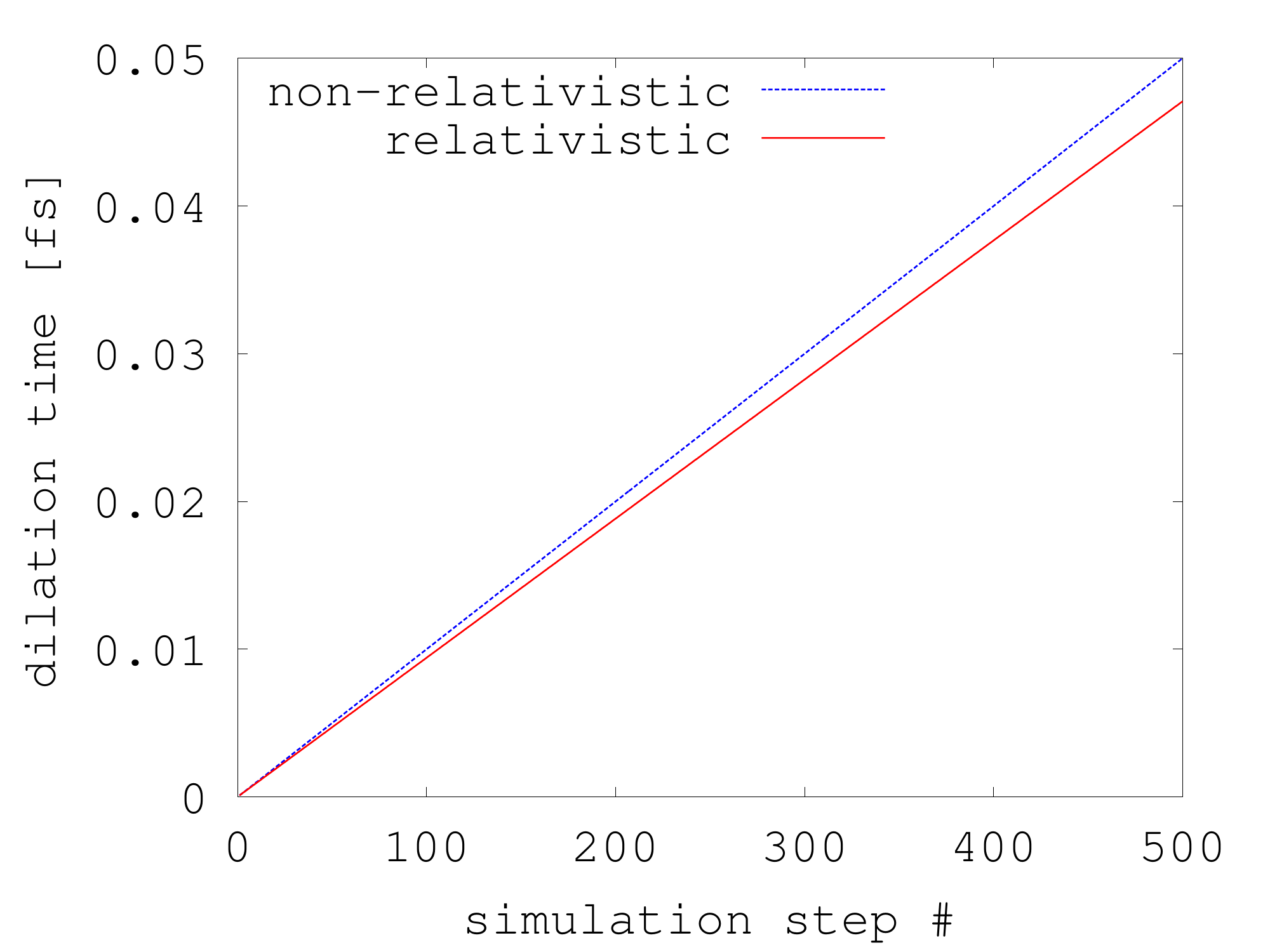}
\end{tabular}
\end{minipage}
\caption{Proper time of an initially Gaussian wave-packet in the non-relativistic case, dashed (blue) curve, and relativistic case, continuous (red) curve, respectively. The phenomenon of time dilation, typical of relativistic theories, is clearly observable.}
\label{time_dilation}
\end{figure}

\begin{figure}[h!]
\centering
\begin{minipage}{0.96\textwidth}
\begin{tabular}{c}
\includegraphics[width=1.0\textwidth]{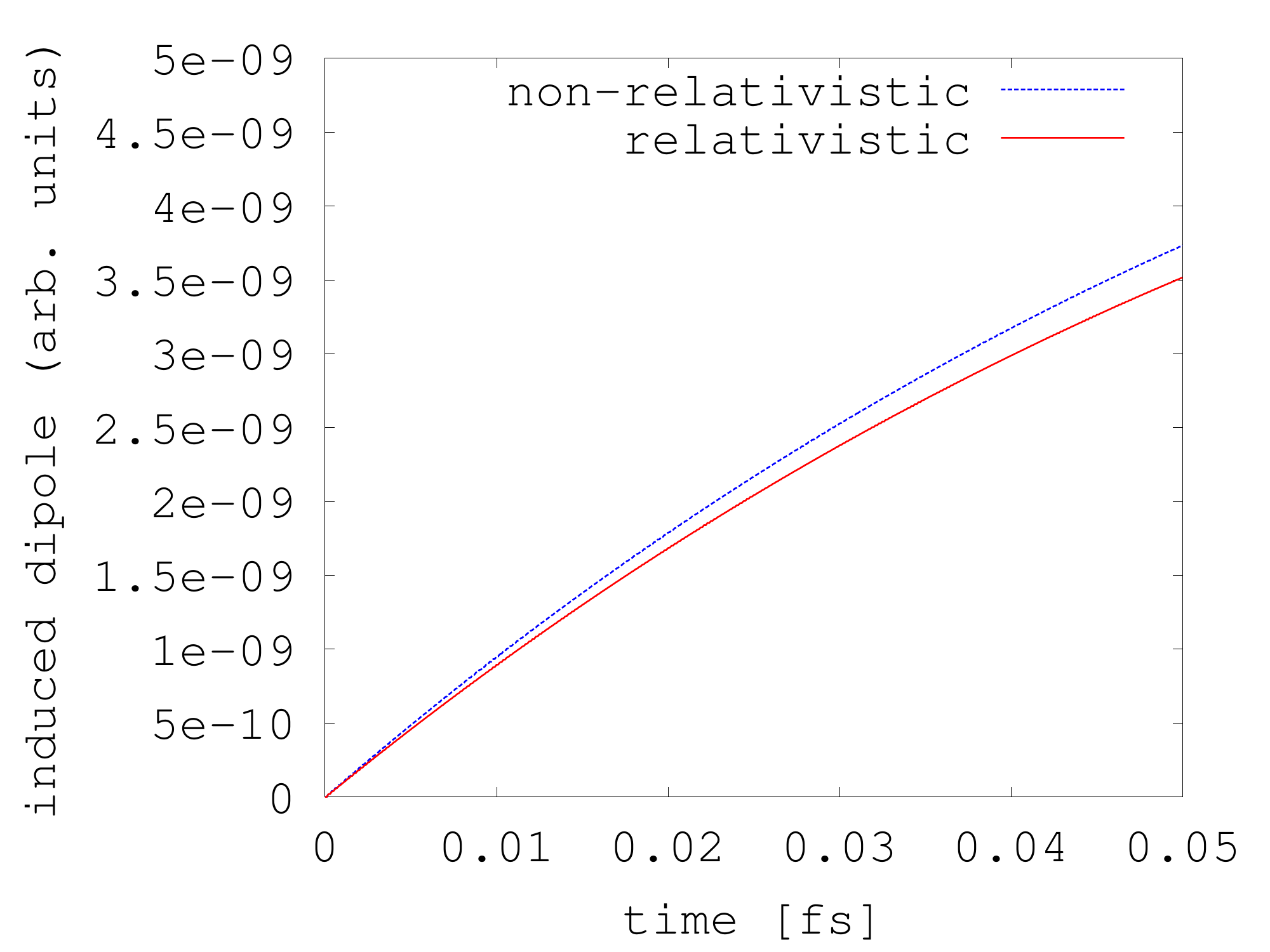}
\end{tabular}
\end{minipage}
\caption{Induced dipole of an initially Gaussian wave-packet in the non-relativistic case, dashed (blue) curve, and relativistic case, continuous (red) curve, respectively. A shift in the energy spectrum, typical of relativistic systems, is clearly visible.}
\label{induced_dipole}
\end{figure}

\section{The classical limit}

In this section we explore the classical limit of our suggested formulation for $\hbar \rightarrow 0$.
To this aim, we distinguish between two cases: dynamics in free space and dynamics in the presence of an external potential.
This distinction allows a natural and simple way to investigate
how the signed particle theory can re-obtain the Newtonian theory.
Although many assumptions are made in this section, we obtain an important point: signed particle creation is prohibited in the classical limit.

\bigskip

{\sl{Free space}}. In the case of free space (absence of any external potential), and in the classical limit $\hbar \rightarrow 0$,
the Heinsenberg principle does not represent a restriction any longer. Therefore, it is possible to start from an initial distribution
function for the signed particles which has the shape of a Dirac $\delta-$function (in other words, one particle with a given position and momentum).
Furthermore, in the classical limit, particles have always a positive sign as it is well-known that classical distribution
functions are positive definite. In this context, thus, a signed particle behaves as a classical Newtonian (field-less) particle
as expected and the classical limit is trivially obtained. It is interesting to note that in the absence of any external
potential the function $\gamma({\bf{x}})$ is identically zero or, in other words, there is no creation of particles.

\bigskip

{\sl{External potential}}. We now discuss the case in the presence of a non-identically-zero external potential $V=V(x)$
in the classical limit $\hbar \rightarrow 0$. For the sake of simplicity, we restrict ourselves to the one-dimensional case,
as the extension to the three-dimensional space is trivial.

One starts by noting that given a potential, and having calculated the
function $\gamma(x)$ from it, the probability for a (positively signed) particle to stay in a given path $x = x(t)$,
during the time interval $[0, t]$, is
$$
 e^{-\int_0^{x=x(t)} \gamma(x(t')) dt'}.
$$
Now, one can show that the Wigner kernel (\ref{wigner-kernel}) reads
$$
 V_W(x; p) = \frac{1}{\pi \hbar^2} \int_{-\infty}^{+\infty} dt' e^{-\frac{ p^2 t'}{m \hbar}} \frac{p}{m}
\left[ V \left( x + \frac{p}{2m} t' \right) - V \left( x - \frac{p}{2m} t'  \right) \right],
$$
under the change of integration variable $x' = \frac{p}{2 m} t'$.
By exploiting the fact that the Wigner kernel is a real function \cite{Wigner}
and supposing that the potential $V=V(x)$ can be expressed in terms of a McLaurin series, one easily obtains
(by means of trivial algebraic manipulations) the following expression for the kernel
\begin{eqnarray}
 V_W(x; p) &=& \frac{2 p}{\pi \hbar^2 m} \sum_{n=0, even}^{+\infty} \int_{-\infty}^{+\infty} dt' \cos(\frac{p^2 t'}{\hbar m})
\frac{V^{(n)}(x)}{n!} \left( \frac{p t'}{2m} \right)^n \nonumber \\
&=& \frac{2 p}{\pi \hbar^2 m} \int_{-\infty}^{+\infty} dt' \cos(\frac{p^2 t'}{\hbar m}) V(x) + \nonumber \\
&+& \frac{2 p}{\pi \hbar^2 m} \int_{-\infty}^{+\infty} dt' \cos(\frac{p^2 t'}{\hbar m}) \frac{1}{2} \frac{\partial^2 V(x)}{\partial x^2} \left( \frac{p t'}{2m} \right)^2 + \nonumber \\
&+& O(\hbar^4), \nonumber
\end{eqnarray}
where the summation runs over {\sl{even}} numbers only and $V^{(n)}(x)$ is the $n-$th order derivative of the potential.
Finally, by neglecting every term down to the second order $O(\hbar^2)$, and by performing the analytical integration of the
only term remaining, one has
\begin{eqnarray}
V_W(x; p) &\approx& \left[ \frac{2 V(x)}{\pi p \hbar} \sin(\frac{p^2 t'}{\hbar m}) \right]_{-\infty}^{+\infty} \nonumber \\
          &=& 2 \lim_{t' \rightarrow +\infty} \frac{2 V(x)}{\pi p \hbar} \sin(\frac{p^2 t'}{\hbar m}) \nonumber \\
          &\approx& \frac{4 V(x)}{\pi p \hbar} \sin(\frac{p^2 T}{\hbar m}), \nonumber
\end{eqnarray}
(for a big enough time $T$) which substituted in (\ref{momentum_integral}) gives the following expression for the function $\gamma(x)$
$$
 \gamma(x) = \lim_{\Delta p \rightarrow 0^+} \sum_{M = -\infty}^{+\infty} \left[ \frac{4 V(x)}{\pi \hbar (M \Delta p)}
\sin(\frac{(M \Delta p)^2 T}{\hbar m})  \right]^+,
$$
with $[ \dots ]^+$ the positive part of the quantity in brackets.
Now, supposing that $V(x)>0$ for any position $x$ (the extension to a general potential is trivial), the condition
for the term in the bracket to be positive reads:
$$
 0 \le M \le \frac{\hbar m \pi}{\Delta p^2 T},
$$
which, in the classical limit $\hbar \rightarrow 0$, becomes
$$ 0 \le M \le 0, $$
or in other words $M=0$. Therefore (in the classical limit) the function $\gamma$ is identically zero $\gamma(x) = 0$,
and creation of pairs of signed particles is prohibited.

Finally, concerning the equations of motion for (positively) signed particles for $\hbar \rightarrow 0$,
one starts from the fact that the Wigner equation reduces to the Vlasov equation \cite{Wigner}. Therefore,
the technique utilized in the previous section to obtain the postulates of our formulation must be applied
to the Vlasov equation. By exploiting the main results of \cite{BMC} (which is an application of the same
technique to the Boltzmann equation) one easily recover the Newtonian equations. This completes the study
of the classical limit.

\section{Simulation of quantum tunnelling}


Tunnelling is a typical quantum effect which cannot be explained in terms of classical mechanics.
In fact, it represents one of the foundational experimental evidence for the need of a quantum theory.
In particular, it shows that material particles can tunnel through potential barriers,
even when the initial particle energy is classically unsufficient.
In this section, we show that such experimental observation can be reproduced
by our suggested formulation of quantum mechanics,
thus proving the validity and the applicability of the approach.

Two benchmark tests are performed (similar to the one presented in \cite{MCMA})
consisting of an initially Gaussian wave-packet with energy equal to $0.08$ eV directed towards
a potential barrier equal to $0.10$ eV (in the first case) and to $0.30$ eV (in the second case).
The simulation of a free wave-packet is also performed for comparison purposes and shown in Fig. \ref{distribution_00}.
In more details, in the first numerical experiment ($0.10$ eV) one expects that part of the wave packet to be
reflected back and part of the packet to tunnel through the barrier (due to its energy
slightly smaller than the barrier itself). In the second experiment, instead, a strong reflection is expected due
to the high energetic value of the barrier ($0.30$ eV).

Every simulation is fully obtained by evolving an initial set of signed particles as prescribed in
the three postulates (introduced above) and augmented by the equations of motion for a signed particle:
\begin{eqnarray}
 {\bf{x}}(t) &=& {\bf{x}}(t-\Delta t) + \frac{{\bf{p}}(t-\Delta t)}{m} \Delta t \nonumber \\
 {\bf{p}}(t) &=& {\bf{p}}(t-\Delta t).
\label{classical-trajectory}
\end{eqnarray}

The results of these two experiments are reported in Figs. \ref{distribution_01} and \ref{distribution_03}
showing the simulation results at times (from left to right, from top to bottom)
$0$ fs, $10$ fs, $20$ fs, $30$ fs, $40$ fs, $50$ fs, $60$ fs and $70$fs respectively.
In particular, one notes in both experiments the appearance of negative values during the evolution, a typical sign
of the presence of dominant quantum effects (in other words tunnelling). Furthermore, it becomes clear
that a certain amount of signed particles tunnels through the barrier (represented
schematically by a red line in the middle of the domain) although this amount varies according to the height of the barrier.
We believe that the fact that quantum tunnelling effects can be
reproduced by applying the operator $\hat{S}$ shows the validity and applicability
of our suggested formulation of quantum mechanics.

\section{Relativistic formulation}

An extension to relativistic quantum mechanics can be delineated taking into account
the classical nature of the particles involved in the new formulation of quantum mechanics.
Indeed, when introducing the second postulate, no equation for the trajectory is specified.
The enunciation is very general and expressed only in terms of {\sl{classical field-less particles}}.
Eventually, the equations are specified for the first time in (\ref{classical-trajectory})
when necessary for implementation purposes, but this is not the only option available.

As a matter of fact, it is possible to define the trajectories of particles involved in postulate II as geodesics,
allowing the evolution of quantum objects in a curved space-time.
Hence, given a metric tensor $g^{\mu \nu}$, describing the geometric structure of space-time, one
can substitute eqns. (\ref{classical-trajectory}) by a relativistic orbit (geodesics):
\begin{equation}
 \frac{d^2 x^{\mu}}{d s^2} = -\Gamma^{\mu}_{\alpha \beta} \frac{d x^{\alpha}}{ds} \frac{d x^{\beta}}{ds},
\label{}
\end{equation}
where $\alpha,\beta,\mu = 0 \dots 3$, $s$ is the proper time, $\Gamma^{\mu}_{\alpha \beta}$ are the Christoffel symbols (a function of $g^{\mu \nu}$),
and $x^0 = t$, $x^1 = x$, $x^2 = y$, $x^3 = z$ ($x,y,z$ spatial coordinates).
In particular, if the metric tensor describes a flat space-time (Minkowski metric), one obtains special relativity.

Figs. \ref{time_dilation} and \ref{induced_dipole} show the results obtained from the simulation of
an initially Gaussian wave-packet moving in a flat space-time with no external potential at a speed comparable to the speed of light
(about $\frac{1}{3} c$).
In particular, Fig. \ref{time_dilation} shows the time dilation computed (as a macroscopic average over all involved particles,
see (\ref{macroscopic-value})) in the non-relativistic (blue dashed curve) and relativistic (red contniuous curve) cases respectively.
Fig. \ref{induced_dipole}, instead, displays the (damped) induced dipole from which one obtains the energy spectrum
of the system through a Fourier transform. It is clear from these pictures how typical relativistic effects,
such as time dilation and energy spectrum shift, are essentially captured by the suggested relativistic extension of our quantum theory.

\section{Conclusions}

In this paper, a new mathematical formulation of quantum mechanics has been suggested in terms of signed Newtonian field-less
particles which can be seen as a generalization and physical interpretation of the time-dependent Wigner Monte Carlo
method \cite{MCMA} extended to a non-discretized phase-space and a non-finite domain. The suggested formulation consists of
a set of three postulates which completely describe the time evolution of quantum systems.
The predictions made by this theory are exactly the same as those made by other standard approaches.
In fact, we have shown that simulating a quantum system by repeatedly applying
the operator $\hat{S}$ is perfectly equivalent to solve the time-dependent Wigner equation and,
consequently, the time-dependent Schr\"{o}dinger equation. In order to show the validity and applicability
of the suggested formulation, we have simulated different systems where quantum tunnelling effects are prominent.
Finally, we have proposed a relativistic extension of the second postulate in terms of signed particles
moving on space-time geodesics and we have shown how typical relativistic effects, such as time dilation and
shift of energy spectrum, are predicted correctly. This particular example underlines how the new formulation
provides a novel significantly easy-to-extend quantum theory.

To conlude, one should note that many aspects of this peculiar and new formulation still remain to be explored.
For instance, it would be interesting to study how the introduction of a finite time step $\Delta t$
affects the simulation of a quantum system (as it may strongly depends on the presence of an external potential).
Likewise, the frequency of annihilation steps to apply in a simulation remains an open problem which
needs to be investigated.
It would be also important to study the mathematical properties (and in particular the convergence)
of the newly defined {\sl{momentum}} integral (\ref{momentum_integral}).
Moreover, it would be useful to investigate the relation between this new formulation and others such as
the ones proposed by D.~Bohm and R.~Feynman (as many similarities can be found already).
It is also clear that the classical limit presented in this work may not be entirely satisfactory
from a mathematical/theoretical perspective as it introduces several strong assumptions which
may need to be investigated in further details.
Another interesting point is to apply this method to bigger and more realistic
quantum systems occurring in real life applications and see how it performs compared to other more standard approaches.
This will be the topic of next-future papers.
Despite all these pending questions, the signed particle formulation remains a new instrument
to explore quantum mechanics, offering a unique and alternative new point of view.

\section{Computational aspects}

The simulator used to  obtain the results presented in this   paper   is a modified version of
Archimedes, the GNU  package for the  simulation of carrier transport in semiconductor devices
\cite{Archimedes}. This code was first released in 2005, and, since then, users have been able
to download the  source    code    under the GNU Public License (GPL). Many features have been
introduced in this package.   In this particular project, our aim has been to develop a full quantum
time-dependent nanodevice simulator including phonon scattering effects.  The code is entirely
developed in C and optimized to get the best performance from the hardware.  It can
run on parallel machines using the OpenMP standard library.  The results of the present version
are posted   on  the   nano-archimedes website,  dedicated to the    simulation of quantum systems.
\cite{nano-archimedes}.

The results have been obtained using the  HPC cluster deployed
at the Institute of Information and Communication Technologies of the Bulgarian Academy of Sciences.
This cluster consists of two racks which contain HP Cluster Platform Express 7000 enclosures
with 36 blades BL 280c with dual Intel Xeon X5560 @ 2.8 Ghz (total 576 cores), 24 GB RAM
per blade. There are 8 storage and management controlling nodes 8 HP DL 380 G6 with dual
Intel X5560 @ 2.8 Ghz and 32 GB RAM. All these servers are interconnected via non-blocking DDR
Infiniband interconnect at 20Gbps line speed. The theoretical peak performance is 3.23 Tflops.

\bigskip

{\bf{Acknowledgements}}.
This paper is dedicated to the beloved memory of Prof. A.M.~Anile. The author would like to
thank Prof. I.~Dimov and Prof. M.~Nedjalkov for the very fruitful conversations, enthusiasm and encouragement.
The author also thanks the anonymous reviewers whose suggestions have significantly helped.
This work has been supported by the project EC~AComIn~(FP7-REGPOT-2012-2013-1).

\end{document}